\documentclass[journal,twoside,web]{ieeecolor}
\usepackage{generic}
\usepackage[noadjust]{cite}
\usepackage{amsmath,amssymb,amsfonts}
\usepackage{graphicx}
\usepackage{textcomp}
\usepackage{cite}
\usepackage{algorithmicx}
\usepackage{textcomp}
\usepackage{subfigure}

\usepackage{extarrows}
\usepackage{accents} 

\usepackage{mathrsfs}

\usepackage{booktabs} 
\usepackage{multirow} 

\usepackage{bm} 

\usepackage{algorithm} 
\usepackage{algpseudocode}

\usepackage[mathscr]{euscript}

\newtheorem{Def}{Definition}
\newtheorem{Rema}{Remark}

\newtheorem{Coro}{Corollary}
\newtheorem{Exam}{Example}

\newtheorem{Propos}{Proposition}
\newtheorem{Lem}{Lemma}
\newenvironment{Pro}{{\it Proof \hspace{-2.7pt}:}\;}{\vspace{5pt} \hfill$\blacksquare$\par} 
\allowdisplaybreaks[1] 

\def\BibTeX{{\rm B\kern-.05em{\sc i\kern-.025em b}\kern-.08em
		T\kern-.1667em\lower.7ex\hbox{E}\kern-.125emX}}
\begin{document}
\title{\huge  Sparse Spectrahedral Shadows for	State Estimation and Reachability Analysis: Set Operations,  Validations and Order Reductions}
\author{Chengrui Wang, Haohao Qiu, Sibo Yao and James Lam$^{*}$, \IEEEmembership{Fellow, IEEE} 
\thanks{*Corresponding author: James Lam (Email:james.lam@hku.hk). }
\thanks{Chengrui Wang, Haohao Qiu and James Lam  are with the Department of Mechanical Engineering, The University of Hong Kong, Pokfulam, Hong Kong. Sibo Yao is with the College of Intelligent Systems Science and Engineering, Harbin Engineering University, Harbin 150001, China }
}
\maketitle

\begin{abstract}
Set representations are  the foundation of various set-based approaches in state estimation, reachability analysis  and fault diagnosis. In this paper, we investigate \textit{spectrahedral shadows}, a class of nonlinear geometric objects previously studied in semidefinite programming and real algebraic geometry. We demonstrate spectrahedral shadows generalize traditional and emerging set representations like ellipsoids, zonotopes, constrained zonotopes and ellipsotopes. Analytical forms of set operations are  provided  including  linear map, linear inverse map, Minkowski sum, intersection, Cartesian product, Minkowski-Firey $\textit{L}_\textit{p}$ sum, convex hull, conic hull and polytopic  map,  all of which are implemented without approximation in  polynomial time. In addition, we develop set validation and order reduction techniques for spectrahedral shadows, thereby establishing spectrahedral shadows  as a set representation applicable to a range of set-based tasks.
\end{abstract}


\section{Introduction}
\label{sec:introduction}
Set-based approaches have been widely used in fields such as state estimation (SE) \cite{schweppe2003, cong2021}, reachability analysis (RA) \cite{halder2020, althoff2014} and fault diagnosis (FD) \cite{xu2019b, scott2016}, since  many of them require handling worst-case  uncertainties (e.g., measurement noise, modeling error, and process disturbance) and establishing reliable boundaries for system behaviors.  As the common foundation of these fields, set representations directly determine the accuracy, the efficiency and the space overhead for implementing  set-based approaches.

\subsection{Existing Convex Set Representations}
The topic of this paper is  convex set representations. In SE, RA  and FD, traditional set representations  include polyhedrons, ellipsoids, zonotopes and support functions \cite{althoff2010, mu2022}, and common set operations involve affine map, Minkowski sum, intersection, convex hull and affine inverse map \cite{scott2016, alamo2005}. Polyhedrons  have    vertex representation (V-polytopes) and  hyperplane representation (H-polyhedrons). The former is closed under  affine map, Minkowski sum and convex hull, and the latter is closed under affine inverse map and intersection. However, the complexity of  conversion  between V-polytopes and H-polyhedrons is NP-hard \cite{dyer1983}, which makes  approaches using polyhedrons extremely time-consuming when dimension exceeds about 5. To the contrary, working with  ellipsoids is highly efficient \cite{scott2016}. Ellipsoids  can  represent practical quadratic boundaries without approximation, such as the bounded energy constraints and the confidence region of Gaussian distribution. Unfortunately, ellipsoids are not closed under  Minkowski sum, convex hull and intersection, and hence need additional ellipsoidal approximation  after these operations, resulting in loss of accuracy. In comparison, zonotopes are a high efficient set-representation closed under affine map and Minkowski sum, but zonotopes  cannot exactly represent   non-centrosymmetric  or nonlinear boundaries.  Support functions can  characterize all non-empty closed convex sets, and have analytical expressions for affine map and Minkowski sum \cite{halder2020}. But in practice, support functions need approximation to implement intersection \cite{le2009} and cannot check emptiness, which restricts the application of support functions on tasks such as  set-membership estimation \cite{alamo2005} and collision avoidance \cite{kousik2023}.

Recently, constrained zonotopes \cite{scott2016, raghuraman2022}, ellipsotopes \cite{kousik2023} and constrained convex generators (CCGs) \cite{silvestre2023, rego2025} have been developed to overcome  the shortcoming of  traditional set representations. Besides the inherent advantages of zonotopes, constrained zonotopes can represent arbitrary convex bounded polytopes \cite[Theorem 1]{scott2016}, and are  closed under intersection and convex hull \cite[Theorem 5]{raghuraman2022}. To handle  mixed polytopic and ellipsoidal uncertainties, ellipsotopes are proposed  based on constrained zonotopes through introducing an indexed set and the Cartesian product of $p$-norm balls.  Ellipsotopes  unify constrained zonotopes and ellipsoids, and inherit the closure properties under all set operations of constrained zonotopes,  except for convex hull \cite[Proposition 8]{kousik2023}. Following ellipsotopes, CCGs further relax the  Cartesian product of $p$-norm balls  as  the  Cartesian product of arbitrary constraints, such that convex hull can be implemented without approximation under an implication  assumption.

\subsection{Overview}

This paper investigates \textit{spectrahedral shadows}, a class of nonlinear convex geometry previously studied by semidefinite programming \cite{helton2009, gouveia2011} and real algebraic geometry \cite{scheiderer2018, bettiol2021}. Different from the previous research, we focus on   necessary techniques to apply spectrahedral shadows on SE, RA and FD. Specifically, our work includes
 \begin{enumerate}
	\item  Section \ref{section: set operations}  provides the nonconservative realization of set operations including   linear map, linear inverse map, Minkowski sum, intersection, Cartesian product, Minkowski-Firey $\textit{L}_\textit{p}$ sum, convex hull, conic hull and polytopic  map. We show that all of them can be implemented within polynomial time.
	\item   Section \ref{section: Set Validations} is the implementation  of  common set validations in SE, RA and FD, including point containment, emptiness check and boundedness check.
	\item   Section \ref{section: relation with existing set representation} gives the nonconservative methods to  convert H-polyhedrons, ellipsoids, zonotopes, constrained zonotopes and ellipsotopes to   spectrahedral shadows.
	\item  Section \ref{section: Complexity Reduction Strategy} discusses the strategies  to reduce the complexity of high order   spectrahedral shadows. Acceleration strategy is also provided    utilizing the sparsity possessed by  spectrahedral shadows  in practice.
\end{enumerate}
Through the above work, spectrahedral shadows are established as a set representation  applicable to set-based SE, RA and FD. Compared with the advanced set representations like ellipsotopes \cite{kousik2023} and CCGs \cite{silvestre2022, rego2025}, spectrahedral shadows are with the following ‌characteristics
 \begin{enumerate}
	\item  (More compact representation) Spectrahedral shadows is concisely represented by  one linear matrix inequality (LMI), instead of the form of ellipsotopes and CCGs that requires the center, generators, linear equalities, $p$-norm balls, index set and even more complex constraints.  Such conciseness‌ makes  spectrahedral shadows representing high order sets  more compactly and efficiently. Our numerical results show that the  space overhead of high-order spectrahedral shadows  is reduced by about 55\%-85\%  compared with the same set represented by  ellipsotopes and CCGs. For high order sets, reducing space overhead means improving the  efficiency.  Because  time consumption of set operations is positively correlated with the input scale of set representations, and using order reduction techniques to reduce space overhead also results in extra time consumption.
	\item  (More supported set operations) Besides the exact set operations realized by  ellipsotopes \cite{kousik2023} and CCGs \cite{silvestre2023},  spectrahedral shadows additionally support affine inverse map,  Minkowski-Firey $\textit{L}_\textit{p}$ sum, conic hull and polytopic  map. Affine inverse map is useful for tasks such as computing the unguaranteed detectable faults set \cite{wang2024}, backward reachability analysis \cite{kvasnica2015, borrelli2017} and the set membership estimation with limited senors (see Section \ref{section: state estimation examples} for example). Thorough Minkowski-Firey $\textit{L}_\textit{p}$,   exact reachable sets can be attained for the task  in \cite{halder2020}. With polytopic  map,  spectrahedral shadows are able   to achieve exact SE and RA on uncertain linear parameter-varying (LPV) systems \cite{xu2019b, silvestre2022}. Moreover, the exact  convex hull can be directly obtained for bounded  spectrahedral shadows, instead of requiring  additional assumptions like CCGs (see \cite[Theorem 1]{silvestre2023}) or outer approximation like ellipsotopes (see \cite[Proposition 8]{kousik2023}).
\end{enumerate}

\section{Preliminaries}

\subsection{Notations}
	The $n\times m$ null matrix and the $n\times m$ matrix full of $1$ are denoted as $\textbf{0}^{n\times m}$ and $\textbf{1}^{n\times m}$, and the  subscript will be omitted for brevity (e.g., $\textbf{0}$) when there is no ambiguity.  The $n$-dimensional identity  matrix is denoted as $I_n$.  	 The notation $\mathbb{N}$ represents the natural numbers. The symbols $\mathbb{R}^n$ and $\mathbb{S}^n$ denote the $n$-dimensional vector space and the symmetric matrix space over  real numbers. For $x, y \in \mathbb{R}^n$, $x_i$  denotes the $i$-th component of $x$,  and $x \geq y$ means $x_i \geq y_i, \, \forall \, 1\leq i \leq n$ (and similar goes for $x \leq y$, $x > y$ and $x < y$). For  $X\in\mathbb{R}^{n\times m_1}$ and $  Y \in\mathbb{R}^{n\times m_2} $, $[X \; Y]$ denotes the matrix concatenation of $X$ and $Y$, $X_{[i,\,j]}$ denotes the element  in the $i$-th row  and $j$-th column of $X$, and $X_{[i,*]}$ and $X_{[*,j]}$ are the $i$-th row  and the $j$-th column of $X$, respectively. 

	For $X \in \mathbb{R}^{n\times n}$, $\mathrm{dvec}(X)$  denotes a  vector such that $\mathrm{dvec}(X)_i = X_{[i,\,i ]}, \, \forall \, 1\leq i \leq n$.
	For $X,Y \in \mathbb{S}^n$, $X \succeq Y$ means $X-Y$ is a positive semidefinite matrix (and similar rules apply for $X \preceq Y$, $X \succ Y$ and $X \prec Y$), and the vector $\,\mathrm{tvec}(X) \in \mathbb{R}^{n(n+1)/2}$  is obtained by  vectorizing the upper (or lower) triangular part of $X$ by column. For $X_1\in \mathbb{R}^{n_1 \times m_1}, X_2\in \mathbb{R}^{n_2 \times m_2},..., X_v\in \mathbb{R}^{n_v \times m_v} $, we  define
	$$ \mathrm{diag}(X_1, \, X_2, ..., X_v )= \left[ \begin{matrix}
		X_1      &	\textbf{0} & \cdots  & \textbf{0} \\
		\textbf{0}  &	X_2 & \cdots  & \textbf{0} \\
		\vdots &	\vdots & \ddots  & \vdots\\
	\textbf{0}  &	\textbf{0} & \cdots  & X_v
	\end{matrix} \right]. $$ 
	 For $X_1, X_2,..., X_v \in \mathbb{R}^{n \times m}$, the convex combination of $X_1, X_2,..., X_v $ is a matrix set   defined as $ \mathrm{conv}\{T_1, \,T_2, \, ... \, ,\, T_v  \} = \{ \sum_{i=1}^{v} \theta_i T_i \! :   \sum_{i=1}^{v} \theta_i =1, \theta_i \geq 0,\, 1 \leq i \leq v\}$.
	
	Consider two sets $X, Y \subset \mathbb{R}^{n}$. The notation $\overline{X}$, $\mathrm{bd}(X)$, $|X|$ and $\mathbb{P}(X)$ denote the  closure, the boundary, the cardinality and the power set of $X$, respectively. The notation $\mathrm{dim}(X)$ denotes the dimension of the affine hull of $X$. The image and preimage of $X$ under the map $f$ are represented by $f\circ X$ and $X\circ f$, respectively. The convex hull of $X$ and $Y$ is defined as $ \mathrm{conv}(X \cup Y) =\{\theta_1x+\theta_2 y \! : x\in X,\, y\in Y,\, \theta_1, \theta_2 \geq 0,\, \theta_1 + \theta_2 =1 \} $. The Minkowski sum,  Cartesian product and  difference of $X$ and $Y$ are denoted by $X \oplus Y = \{x+y: x\in X,\, y\in Y \}$, $X \times Y = \{[x,\, y]^{T} \! : \, x\in X,\, y\in Y \}$ and $X \backslash Y = \{x: x\in X,\, x\notin Y \}$, respectively. The unit $p$-norm ball in $\mathbb{R}^n$ is denoted by $\mathcal{B}_{p}^{n} = \{x  \in \mathbb{R}^n\! : \left\| x\right\|_p \leq1 \}$. The volume of  $X$ is denoted as $\mathrm{vol}(X)$, and the volume ratio of  $X$ and $ Y $ is defined as $\Delta V =  (\frac{\mathrm{vol}(X)}{\mathrm{vol}(Y)})^{\frac{1}{n}}$. \vspace{1pt}

\subsection{Set Representations}

  This section introduces   traditional and emerging convex set representations \cite{combastel2003,blanchini2015,scott2016,kousik2023} used in SE, RA and FD.
  
  \subsubsection{H-polyhedron}  \textit{H-polyhedrons} are a collection of sets represented by the intersection of finite  half-spaces, i.e., $\mathcal{P}(A, b) = \{x \in \mathbb{R}^{n}\! : Ax\leq b\}$, where $A\in \mathbb{R}^{n_c\times n}$ and $b \in \mathbb{R}^{n_c}$  define the hyperplanes. \vspace{1pt}
  
  \subsubsection{Ellipsoid}  \textit{Ellipsoids} are a collection of sets represented by $\mathcal{E}(c,\, Q) = \{x \in \mathbb{R}^{n}: (x-c)^TQ^{-1}(x-c) \leq 1,\, Q\succ 0 \}$, where $Q\in \mathbb{S}^{ n}$ and $c \in \mathbb{R}^{n}$ is called the \textit{center}. \vspace{1pt}
  
  \subsubsection{Zonotope} \textit{Zonotopes} are a  collection of centrosymmetric polytope represented by $\mathcal{Z}(c,\, G) = \{ x \in \mathbb{R}^{n}: x=c+G\xi,\, \left\| \xi \right\|_\infty \leq 1 \}$, where  $c \in \mathbb{R}^{n}$   and $G\in \mathbb{R}^{ n \times n_g}$ are called the  \textit{center} and  \textit{generator matrix}. \vspace{1pt}

  \subsubsection{Constrained Zonotope} \textit{Constrained zonotopes} are a collection of polytopes in zonotopic form. A constrained zonotope is represented by $\mathcal{CZ}(c,\, G, \, A, \, b) = \{ x \in \mathbb{R}^{n}:  x=c+G\xi,\,  \left\| \xi \right\|_\infty \leq 1, \, Ax =b \}$, where $A \in \mathbb{R}^{n_c \times n_g}$ and $b \in\mathbb{R}^{n_c} $ together define the constraints, and $c \in \mathbb{R}^{n}$   and $G\in \mathbb{R}^{ n \times n_g}$ are consistent with zonotopes. \vspace{1pt}
  
  \subsubsection{Ellipsotope}  \textit{Ellipsotopes} are the unification of ellipsoids and constrained zonotopes. An ellipsotope is represented by  $\mathcal{E}_p(c,\, G, \, A, \, b, \, \mathscr{J}) = \{ x \in \mathbb{R}^{n}:  x=c+G\xi,\,  \left\| \xi\langle J \rangle \right\|_p \leq 1, \, \forall \, J \in \mathscr{J}, \, Ax =b \}$, where $p \geq 1$, $ \mathscr{J}= \{J_1, J_2, ..., J_{|\mathscr{J}|}  \} \subset \mathbb{P}(\mathbb{N})$ is called a \textit{indexed set},  each $\xi\langle J \rangle$ is a part of the vector $\xi$ indexed by $ \mathscr{J}$, and $c \in \mathbb{R}^{n}$, $G\in \mathbb{R}^{ n \times n_g}$, $A \in \mathbb{R}^{n_c \times n_g}$ and $b \in\mathbb{R}^{n_c} $ are consistent with constrained zonotopes.

\subsection{Computational Complexities Analysis} \label{section: Computational Complexities Analysis}
   For the convenience of subsequent analysis,  we introduce some conventions  and  known results about computational complexities used in this paper. We use big O notation (i.e., $\mathcal{O}(\cdot)$) to analyze the worst-case of asymptotic complexities and use $\omega$ to denote the exponent of matrix  multiplication.  Currently, the bound of $\omega$ is $\omega \leq 2.372$ \cite{williams2024}.

    Consistent with   \cite{kousik2023} and \cite{kochdumper2020}, we consider  the time complexity of all matrix operations  except for matrix initialization and matrix concatenation. Specifically, if we consider the complexity of arithmetic for each element  as  $\mathcal{O}(1)$, then we can conclude the following complexity of  matrix operations:
   \begin{enumerate}
   	\item   For $X, Y \in \mathbb{R}^{m \times n}$, $X + Y $ is $\mathcal{O}(mn)$.
   	\item   For $X \in \mathbb{R}^{m \times n}, Y \in \mathbb{R}^{n \times l}$, $XY $ is $\mathcal{O}(mnl)$.
    \item   For $X \in \mathbb{R}^{n \times n}$, $X^{-1} $ is $\mathcal{O}(n^\omega)$.
     \item For $X \in \mathbb{R}^{m \times n}$, the singular value decomposition (SVD) is $\mathcal{O}(\tilde{m}^2 \tilde{n})$, where $\tilde{m} = \max\{m,\,n\}$ and $\tilde{n} = \min\{m,\,n\}$.
   \end{enumerate}
    Moreover, consider the following dual form of  semidefinite programming (SDP) problem:
   \begin{equation} \label{SDP problem}
  	\begin{aligned}
  		\min_{x} \; b^T x \quad s.t., \varLambda  + \sum_{i=1}^{n} x_i A_i\succeq 0,
  	\end{aligned}
  \end{equation}
  where $b,x \in \mathbb{R}^{n}$ and $\varLambda, A_1,...,A_n\in \mathbb{S}^{s} $. For numerically solving the   SDP problem \eqref{SDP problem} using interior point method with given accuracy, the state-of-the-art algorithm runs  in $\mathcal{O}(\sqrt{s}(ns^2+n^\omega+s^\omega))$ time \cite[Theorem I.2]{jiang2020}.

\section{Spectrahedral Shadows} \label{section: spectrahedral shadows}

We first introduce the definition of \textit{spectrahedral shadows}.\vspace{2pt}

\begin{Def}
	A set $\mathcal{S}\subset \mathbb{R}^{n} $ is a \textit{spectrahedral shadow} if there exist $\varLambda, A_1,A_2,...,A_n,B_1,B_2,...,B_m \in \mathbb{S}^{s} $ such that
	   \begin{equation} \label{def of spectrahedral shadows}
		\begin{aligned}
	\hspace{-8pt}	\mathcal{S} = \{ x\in \mathbb{R} ^n \! : \, \exists \, y\in \mathbb{R} ^m, \, \varLambda +\sum_{i=1}^n{x_iA_i}+\sum_{i=1}^m{y_iB_i}\succeq 0  \},\hspace{-1pt}
		\end{aligned}
	\end{equation}
	where $s$ and $m$ are the \textit{size} and \textit{lifted dimension} of spectrahedral shadows, respectively. For brevity, a spectrahedral shadow in the form of \eqref{def of spectrahedral shadows} is abbreviated as $ \langle \varLambda , \, \{ A_i \} _{i=1}^{n},\, \{ B_i \} _{i=1}^{m} \rangle  $. \vspace{3pt}
\end{Def}
   
Spectrahedral shadows have also been called  \textit{SDP representable sets} \cite{helton2010} or \textit{projections of spectrahedrons} \cite{gouveia2011} for the research of semidefinite representability. Moreover,  spectrahedral shadows with zero lifted dimension, i.e.,  $\langle  \varLambda , \, \{ A_i \} _{i=1}^{n} \rangle $,	are called  \textit{spectrahedrons}.

\begin{Rema} \label{sparse form remark}
	 When applied to SE, RA and FD,  spectrahedral shadows  have sparse structures (see Section \ref{section: Acceleration using  Sparsity }). To store $ \langle \varLambda , \, \{ A_i \} _{i=1}^{n},\, \{ B_i \} _{i=1}^{m} \rangle  $ in practice, we only need to store $n$, $s$, $m$ and an $s\times s(1+n+m)$ sparse matrix by stacking the upper triangular part of $ \varLambda , \, \{ A_i \} _{i=1}^{n}$ and $ \{ B_i \} _{i=1}^{m}$ by column. 
\end{Rema}

 Lemma \ref{lemma 1} introduces some known conclusions  used in this paper.

\begin{Lem}[{\cite{helton2009}}] \label{lemma 1}
	There are the following properties about spectrahedral shadows:
	\begin{enumerate}
		\item   A spectrahedral shadow is not necessarily a bounded set. 
		\item   A spectrahedral shadow is not necessarily a closed set. \;\;
		\item   Consider a set $X = \{ x\in \mathbb{R}^n\!: f_i(x) \leq0 , \, i =1,2,$ $..,m \}$, where each $f_i(x)$ is a  polynomial and $X$ has the nonempty interior. If each $f_i(x)$ is quasi-convex on $X$,   $X$ can be represented as a  spectrahedral shadow.
	\end{enumerate}
	\vspace{2pt}
\end{Lem}

\begin{figure}[!t]
	\centering
	\subfigure[An unbounded   2-dimensional spectrahedral shadow]{
		\hspace{-8pt} \includegraphics[scale=0.255]{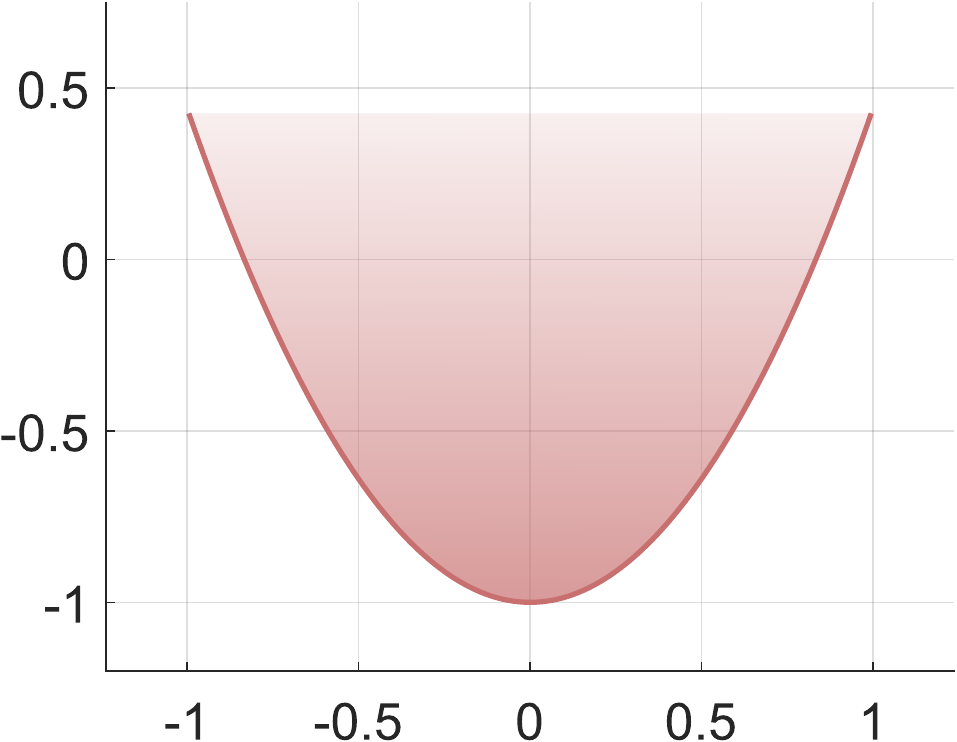}
	}
	\subfigure[A bounded 3-dimensional spectrahedron]{
		\hspace{-8pt} \includegraphics[scale=0.16]{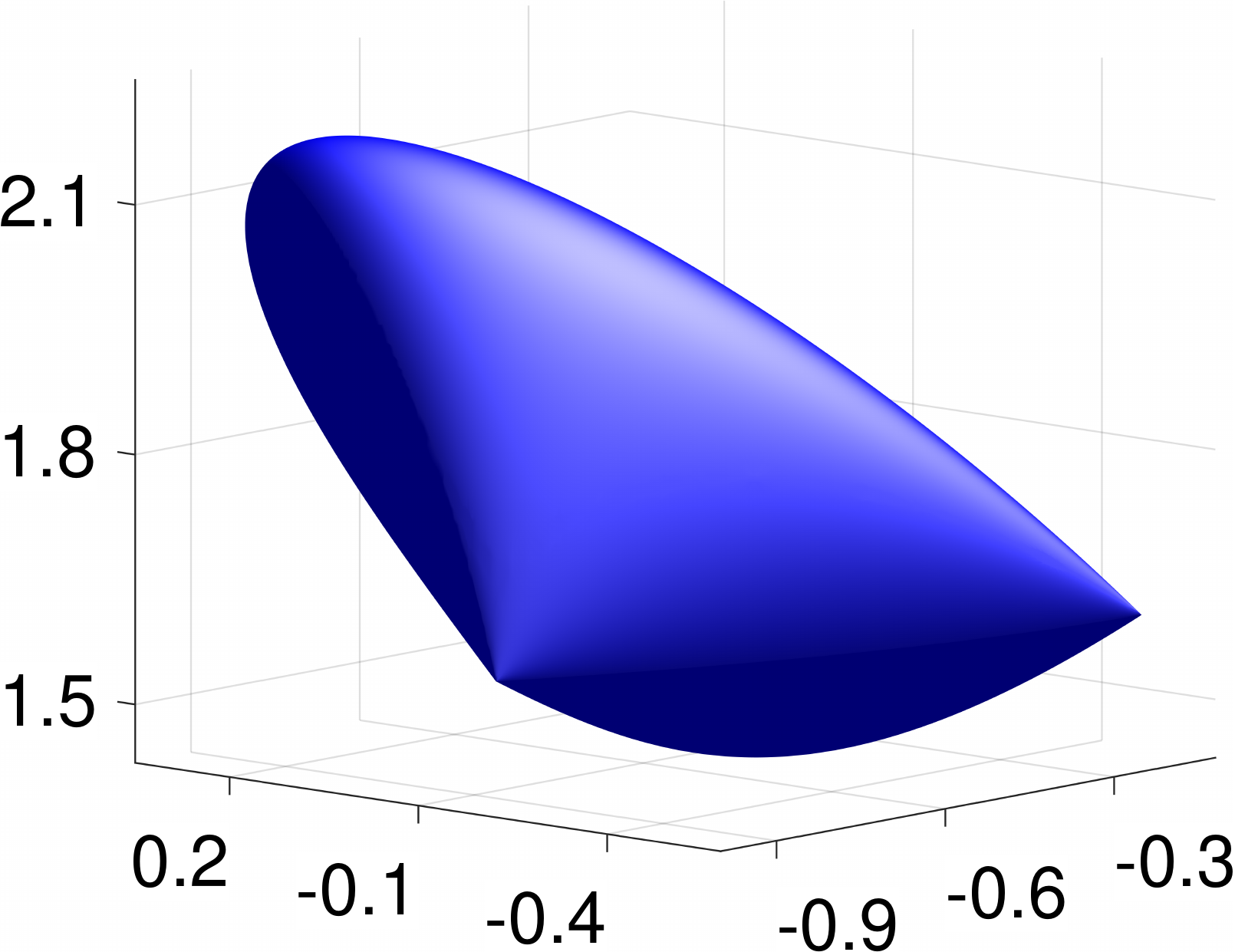}
	}
	\vspace{-4pt}
	\caption{ Spectrahedral shadows given in Example \ref{example of noncompact}.} \label{fig1}
\end{figure}

\begin{Exam} \label{example of noncompact}
  We give a few  examples to illustrate Lemma \ref{lemma 1}. For the point 1), the spectrahedral shadow $\langle  \varLambda , \, \{ A_i \} _{i=1}^{2} \rangle $  with $ \varLambda= I_2 $, $ A_1= 1.2\times(\textbf{1}-I_2) $ and $A_2= \mathrm{diag}(1,\, 0) $ is the region restricted by the parabola $ \{ x\in \mathbb{R}^2\!: 1-1.44x_1^2+x_2=0 \}$ (see Fig. \ref{fig1}(a)), and hence is an unbounded set. For the point 2), the spectrahedral shadow $ \langle \varLambda , \, \{ A_i \} _{i=1}^{1},\, \{ B_i \} _{i=1}^{1} \rangle  $ with $ \varLambda= \textbf{1}-I_2 $, $ A_1= \mathrm{diag}(1,\, 0) $ and $ B_1= \mathrm{diag}(0,\, 1) $  is the open interval $(0, \,\infty)$. For the point 3), all polynomials in the set $ \{ x\in \mathbb{R}^3\!: -x_1 \le 1, \; x_{1}^{2}+x_{2}^{2}+x_{3}^{2}-x_1x_2+x_1x_3-2x_2x_3-x_1+3x_2-3x_3\le-2, \,\; x_{1}^{2}+2x_{2}^{2}+2x_1x_2+2x_1x_3+2x_2x_3-2x_1-2x_2 \le 0,\, \; x_{1}^{2}+x_{2}^{2}+x_1x_2-x_1x_3+3x_1+x_2-x_3\le -2 , \, \; 3x_{1}^{3}-2x_{2}^{3}+x_{3}^{3}+4x_{1}^{2}x_2+8x_{1}^{2}x_3-4x_1x_{2}^{2}+4x_1x_{3}^{2}+x_2x_{3}^{2}+8x_1x_2x_3-5x_{1}^{2}-2x_{2}^{2}-x_{3}^{2}-8x_1x_2+2x_2x_3-4x_1-4x_2\le 0 \}$ are convex (a special case of quasi-convex) on this set. This set can be   characterized  by a spectrahedron concisely, i.e., $\langle  \varLambda , \, \{ A_i \} _{i=1}^{2} \rangle $ with
  \begin{equation} \notag
  	\begin{aligned}
  		\varLambda &= \! \left[ \begin{matrix}
  			\,1    &	-1 & 1 \,  \\
  			\,*  &	-1 & 1 \,\\
  			\,* &	* & -1 \,   
  		\end{matrix} \right]\!\! , \,
  		A_1 = \! \left[ \begin{matrix}
  			\,1     &	1 & 1 \,  \\
  			\,*  &	0 & -1 \,\\
  			\,* &	* & 0 \,   
  		\end{matrix} \right]\!\! ,  \\ \noalign{ \vskip 3pt}
  		A_2 &= \! \left[ \begin{matrix}
  			\,0     &	-1 & 1 \,  \\
  			\,*  &	-1 & -1 \,\\
  			\,* &	* & 1 \,   
  		\end{matrix} \right]\!\! ,		
  		A_3 = \! \left[ \begin{matrix}
  			\,0     &	1 & 0 \,  \\
  			\,*  &	1 & -1 \,\\
  			\,* &	* & 1 \,   
  		\end{matrix} \right]\!\! ,
  	\end{aligned}
  \end{equation}
    where the figure   is shown in Fig. \ref{fig1}(b).
   \vspace{2pt}
\end{Exam}

 \vspace{2pt}

	Section \ref{section: relation with existing set representation}  provides methods to convert existing common set representations  into  spectrahedral shadows. For converting the sets represented by  convex polynomial inequalities to spectrahedral shadows, there is an available construction techniques known as the moment  method, where the details can be found in \cite{lasserre2009, helton2012}.

\subsection{Set Operations} \label{section: set operations}
 Spectrahedral shadows are known to be closed under  operations including affine map, Minkowski sum, intersection, Cartesian product  and convex hull \cite{nemirovski2006, gouveia2011, scheiderer2018}. However, except for the intersection and convex hull,  general expressions have not been systematically seen in the literature\footnote{\cite[Section 4.1.1]{nemirovski2006} asserted that the mentioned set operations  are fully algorithmic, but no  expressions are given except for the intersection (see Lemma \ref{lemma of intersection}) and a special  case of conic hull (see Remark \ref{remark of conic hull}).}. Moreover, in fields such as  set-based SE, RA and FD, our concern is  how to perform these set operations in a unified and iterable manner, and more importantly, whether such operations can be performed in  a reasonable time (e.g., polynomial time).   These issues are  what we aim at addressing in this section.

\subsubsection{Affine Map}
Since an affine map comprises a translation and a linear map, we first give the expression of translation and then give that of linear map for spectrahedral shadows.\vspace{3pt}

	\begin{Lem}[Translation] \label{propos of translation}
	For any $b\in \mathbb{R}^n$ and nonempty $ \mathcal{S} =  \langle \varLambda,\,$ $ \{ A_i \} _{i=1}^{n}, \, \{ B_i \} _{i=1}^{m} \rangle $ with size $s$, we have
	\begin{equation} \label{formula of translation}
		\begin{aligned}
		\mathcal{S} + b = \langle \varLambda -  \sum_{i=1}^{n} b_i A_i,\, \{ A_i \} _{i=1}^{n}, \, \{ B_i \} _{i=1}^{m} \rangle.
		\end{aligned}
	\end{equation} 
	The  computational complexity   is  $\mathcal{O}(ns^2)$. \vspace{3pt}
    \end{Lem}
    
    \begin{Pro}
    For arbitrary  $x\in \mathcal{S}$ and $z = x+b$, we have $z-b \in \mathcal{S}$. It follows from Definition \ref{def of spectrahedral shadows} that  $\exists \, y\in \mathbb{R}^m,\,\varLambda +\sum_{i=1}^n{(z_i-b_i)A_i}+\sum_{i=1}^m{y_iB_i}\succeq 0 $, thereby leading to \eqref{formula of translation}.\vspace{1pt}
    
    \textit{Complexity:} Computing $\varLambda -  \sum_{i=1}^{n} b_i A_i$ needs  $\mathcal{O}(ns^2)$  multiplications and  $\mathcal{O}(ns^2)$ matrix subtractions.  By omitting  the constant coefficients, the overall complexity is   $\mathcal{O}(ns^2)$.
    \end{Pro}
    
    Then  we introduce Lemma \ref{lemma of linear invertible map}   as the foundation to establish the general expression of linear map. \vspace{3pt}
    
    \begin{Lem}[Invertible linear  map] \label{lemma of linear invertible map}
    	For any invertible matrix $T\in \mathbb{R}^{n\times n}$ and nonempty $ \mathcal{S} =  \langle \varLambda,\,$ $ \{ A_i \} _{i=1}^{n}, \, \{ B_i \} _{i=1}^{m} \rangle $ with size $s$, we have
    	\begin{equation} \label{formula of linear invertible map}
    		\begin{aligned}
    		T \circ	\mathcal{S} = \langle \varLambda,\, \{ \sum_{i=1}^{n} T^{-1}_{[i,\, j]} A_i \} _{j=1}^{n}, \, \{ B_i \} _{i=1}^{m} \rangle.
    		\end{aligned}
    	\end{equation} 
    	The  computational complexity   is  $\mathcal{O}(n^{\omega}+n^2s^2)$. \vspace{3pt}
    \end{Lem}
    
     \begin{Pro}
    	For any  $x\in \mathcal{S}$ and $z = Tx$, we have $T^{-1}z \in \mathcal{S}$ as $T$ is invertible, which is equivalent to   $ \exists\, y\in \mathbb{R}^m,\,\varLambda +\sum_{i=1}^n{x_iT^{-1}_{[i,\, j]}A_i}+\sum_{i=1}^m{y_iB_i}\succeq 0 $, thereby leading to \eqref{formula of linear invertible map}.\vspace{1pt}
    	
    	\textit{Complexity:}  The complexity to obtain $T^{-1}$ is $\mathcal{O}(n^{\omega})$, where the constant $\omega$ has been defined in Section \ref{section: Computational Complexities Analysis}. Computing $\sum_{i=1}^{n} T^{-1}_{[i,\, j]} A_i$ \vspace{1pt} for all $1\leq j \leq n$ requires    $\mathcal{O}(n^2s^2)$  multiplications and  $\mathcal{O}(n(n-1)s^2)$ matrix additions.  Thus, the overall complexity is   $\mathcal{O}(n^{\omega}+n^2s^2)$.
    \end{Pro}

       \begin{Propos}[Linear  map] \label{propos of linear  map}
    	For arbitrary   $T\in \mathbb{R}^{l\times n}$ and nonempty $ \mathcal{S} =  \langle \varLambda,\,$ $ \{ A_i \} _{i=1}^{n}, \, \{ B_i \} _{i=1}^{m} \rangle $ with size $s$, we have
    	\begin{equation} \label{formula of linear  map}
    		\begin{aligned}
    			T \circ	\mathcal{S} = \langle \varLambda^t,\, \{ \sum_{i=1}^{n} U^{T}_{[i,\, j]} A_i^t \} _{j=1}^{l}, \, \{ B_i^t \}_{i=1}^{m_t} \rangle
    		\end{aligned}
    	\end{equation} 
    	  with
    		\begin{flalign}\notag
    		&\ \varLambda^t = \mathrm{diag}( 	\varLambda, \, \textbf{0}),&\\
    		&\ A_i^t = \begin{cases}
    			\mathrm{diag}(  \sum_{j=1}^{n} \! D_{[i,\, i]}^{-1} V^T_{[i,\, j]}A_j ,  \, \textbf{0}), \;\;\; 1\leq i \leq r_t \vspace{1pt} \\
    			\mathrm{diag}(\textbf{0} , \, \varGamma^{i-r_t}),  \;\;\; r_t+1 \leq i \leq l\\
    		\end{cases} \hspace{-10pt},& \notag
    		\end{flalign}
    	   		\begin{equation} \notag
    		\begin{aligned}
    		&	{B}_i^t =  \begin{cases}
    				\mathrm{diag}(  B_i ,  \, \textbf{0}), \;\;\, 1\leq i \leq m \vspace{1pt} \\
    				\mathrm{diag}( \sum_{j=1}^{n} \! V^T_{[i-m+r_t, \,j]} A_j, \, \textbf{0} ),  \;\;\, m+1 \leq i \leq m_t\\
    			\end{cases} \hspace{-12pt}, \\
    				& \varGamma^i_{[j, k]} = \begin{cases}
    				1, \;\;\; j=k=i \\
    				-1,  \;\;\; j=k=l-r_t+i\\
    				0 ,  \;\;\; \text{otherwise}  
    			\end{cases} 1 \leq i \leq l-r_t, 
    		\end{aligned}
    	\end{equation}
    	where $\varLambda^t, A_i^t, {B}_i^t  \in \mathbb{S}^{s+2(l-r_t)}$, $\varGamma^i  \in \mathbb{S}^{2(l-r_t)} $, $m_t = m+n-r_t$, $r_t = \mathrm{rank}(T)$, and $T = UDV^T$ is the  SVD such that $D_{[i,\,i]}$ is the $i$-th largest singular value of $T$. The  computational complexity   is   $\mathcal{O}(\tilde{l}^4+ \tilde{l}^3s + \tilde{l}^2s^2 )$, where $\tilde{l} = \max\{l, n\}$. \vspace{3pt}
    \end{Propos}
    
    \begin{Pro}
    	Let $\mathcal{S}_{t}^1 =  V^T \circ \mathcal{S}$ and $\mathcal{S}_{t}^2 =  D \circ \mathcal{S}_{t}^1$. Then we have  $T \circ	\mathcal{S} = U \circ \mathcal{S}_{t}^2$. Since $V^T$ is an orthogonal matrix,  it yields $\mathcal{S}_{t}^1 =  \langle \varLambda,\, \{ \sum_{i=1}^{n} V_{[i,\, j]} A_i \} _{i=1}^{n}, \, \{ B_i \} _{i=1}^{m} \rangle$ by  Lemma  \ref{lemma of linear invertible map}. To formulate $\mathcal{S}_{t}^2 =  D \circ \mathcal{S}_{t}^1$, assume that $z = Dx $ and $x\in \mathcal{S}_{t}^1$. According to Definition \ref{def of spectrahedral shadows}, $x\in \mathcal{S}_{t}^1$ is equivalent to
    	\begin{equation} \label{linear map tmp1}
    		\begin{aligned}
   \exists\, y\in \mathbb{R}^m,\, 	\varLambda +\sum_{i=1}^n{x_i\sum_{j=1}^n{V_{\left[ i,j \right]}^TA_j}}+\sum_{i=1}^m{y_iB^i}\succeq 0.
    		\end{aligned}
    	\end{equation}
    	Since $D$ is a diagonal matrix, $z = Dx$ can be rewritten as
    	\begin{equation} \label{linear map tmp2}
    		\begin{aligned}
    		 z_i=\begin{cases}
    			D_{[i,\,i]} x_i, \;\;\,    1\le i\le r_t\\
    			0,   \;\;\;     r_t+1\le i\le l\\
    		\end{cases}\hspace{-10pt}. 
    		\end{aligned}
    	\end{equation}
    	Note that $	D_{[i,\,i]}>0, \forall \,   1\le i\le r_t$. Substituting \eqref{linear map tmp2} into \eqref{linear map tmp1},  $z = Dx$ and $ x\in \mathcal{S}_{t}^1$ iff  $\sum_{i=r_t+1}^l{z_i\varGamma^{i-r_t}} \succeq 0$ and
    	\begin{equation} \notag
    		\begin{aligned}
    		   \exists\, y\in \mathbb{R}^m,\, 	\varLambda +\sum_{i=1}^{r_t}{z_i\sum_{j=1}^n{D_{[i,\,i]}^{-1}V_{\left[ i,j \right]}^TA_j}}+\sum_{i=1}^m{y_iB_i}\succeq 0,
    		\end{aligned}
    	\end{equation}
    	which yields $\mathcal{S}_{t}^2 =  D \circ \mathcal{S}_{t}^1 = \{Dx: x\in  \mathcal{S}_{t}^1 \} = \langle \varLambda^t,\,  \{ A_i^t \} _{i=1}^{l}, $ $\, \{ B_i^t \}_{i=1}^{m_t} \rangle$. According to Lemma \ref{lemma of linear invertible map}, we  obtain \eqref{formula of linear  map} from  $T \circ	\mathcal{S} = U \circ \mathcal{S}_{t}^2$  and $\mathcal{S}_{t}^2=  \langle \varLambda^t,\,  \{ A_i^t \} _{i=1}^{l}, \, \{ B_i^t \}_{i=1}^{m_t} \rangle$. \vspace{1pt}
    	
    	 \textit{Complexity:} Conducting SVD for $T$ has  $\mathcal{O}(\tilde{l}^2\tilde{n})$ complexity, where $\tilde{n} = \min\{l, n\}$. The construction of $A_i^t$ for all $1 \leq i \leq l$  requires   $\mathcal{O}(r_t(ns^2+1))$  multiplications and $\mathcal{O}((n-1)s^2)$ matrix additions, and hence has $\mathcal{O}(r_tns^2)$ complexity. Similarly, constructing $B_i^t$ for all $1 \leq i \leq l$ has $\mathcal{O}((n-r_t)ns^2)$ complexity. According to Lemma \ref{lemma of linear invertible map}, computing \eqref{formula of linear  map} based on $ \varLambda_t, \, A_i^t$ and $B_i^t$ has $\mathcal{O}(l^2(s+2(l-r_t))^2)$. Since the complexity  of all other operations is a lower-order term, the overall complexity is $\mathcal{O}(\tilde{l}^2\tilde{n} +  n^2s^2 + l^2(s+2(l-r_t))^2)$, which is further approximated as   $\mathcal{O}(\tilde{l}^4+ \tilde{l}^3s + \tilde{l}^2s^2 )$ using  $\tilde{l} = \max\{l, n\}$, $\tilde{n} \leq \tilde{l}$ and $ l-r_t \leq \tilde{l}$.
    \end{Pro}

\subsubsection{Affine inverse map}
 The affine inverse map is useful for set-membership estimation \cite{alamo2005} and backward reachability analysis \cite{kvasnica2015, borrelli2017}. We provide the expression of the linear inverse map in Proposition \ref{propos of linear inverse map}, which leads to the affine inverse map together with Lemma \ref{propos of translation}.	\vspace{3pt}

\begin{Propos}[Linear inverse map] \label{propos of linear inverse map}
	For arbitrary   $T\in \mathbb{R}^{n\times l}$ and nonempty $ \mathcal{S} =  \langle \varLambda,\,$ $ \{ A_i \} _{i=1}^{n}, \, \{ B_i \} _{i=1}^{m} \rangle $ with size $s$, we have
	\begin{equation} \label{formula of linear inverse map}
		\begin{aligned}
				\mathcal{S} \circ T  = \langle \varLambda,\, \{ \sum_{i=1}^{n} T_{[i,\, j]} A_i \} _{j=1}^{l}, \, \{ B_i \}_{i=1}^{m} \rangle 
		\end{aligned}
	\end{equation} 
 The  computational complexity   is    $\mathcal{O}(lns^2)$. \vspace{3pt}
\end{Propos}

	\begin{Pro} 
	 For $x = Tz$, we have $x_i = \sum_{j=1}^l{T_{[i,\, j]}z_j},\, \forall \, 1\leq i \leq n$. Thus,  $x\in \mathcal{S}, \, x = Tz$ iff   $ \exists\, y\in \mathbb{R}^m,\,\varLambda +\sum_{i=1}^n{\sum_{j=1}^l{T_{[i,\, j]}z_jA_i}}+\sum_{i=1}^m{y_iB_i}\succeq 0 $, which yields \eqref{formula of linear inverse map}.\vspace{1pt}
	
	\textit{Complexity:}  Computing $\sum_{i=1}^{n} T_{[i,\, j]} A_i$  for all $1\leq j \leq l$ requires   $\mathcal{O}(lns^2)$  multiplications and  $\mathcal{O}(l(n-1)s^2)$ matrix additions.  The overall complexity is   $\mathcal{O}(lns^2)$.
	\end{Pro}
 
Moreover, Proposition 3 indicates that a set representation cannot be closed under linear inverse map unless this set representation can characterize unbounded sets.	\vspace{3pt}

\begin{Propos} \label{unbounded propos}
	For a set representation $\mathcal{R}$,  there exists an unbounded set  $X \in \mathcal{R}$ if $\mathcal{R}$ is closed under linear  \vspace{2pt} inverse maps. 
\end{Propos}

\begin{Pro} 
	 Without loss of generality, let the singleton set $\{\textbf{0} \} \subset \mathbb{R}^{m} $ be represented by $\mathcal{R}$. Consider  a linear map $T\! : \mathbb{R}^{n} \rightarrow \mathbb{R}^{m}$ that admits a nontrivial null space $\mathrm{Null}(T)$. By definition, the preimage of  $\{\textbf{0} \}$ under $T$ is $\mathrm{Null}(T)$. Since $\mathcal{R}$ is closed under linear  inverse maps, we have $\mathrm{Null}(T) \in \mathcal{R}$.
\end{Pro}

\begin{Coro} \label{unbounded corollary}
	Ellipsoids, zonotopes, constrained zonotopes, V-polytopes and ellipsotopes are not closed under linear inverse maps.  \vspace{2pt}
\end{Coro}

\begin{Pro}
	Corollary \ref{unbounded corollary} directly follows from Proposition \ref{unbounded propos} since all set representations mentioned above can only represent bounded sets.   
\end{Pro}

 \begin{Rema}
	It should be noted that Corollary \ref{unbounded corollary} only asserts that the mentioned set representations are not closed under general  linear inverse maps, but these set representations can be closed with extra  assumptions (e.g., the linear inverse map $T$ is invertible \cite[Section 4.4.11]{borrelli2017}).	\vspace{3pt}
\end{Rema}

\subsubsection{Minkowski sum}
    Then Proposition \ref{propos of Minkowski sum} shows how to compute the Minkowski sum of spectrahedral shadows.  
	\begin{Propos}[Minkowski sum] \label{propos of Minkowski sum}
		Given  $ \mathcal{S}_1 =  \langle \varLambda,$  $\, \{ A_i \} _{i=1}^{n},\, \{ B_i \} _{i=1}^{m_1} \rangle $ with size $s_1$ and  $ \mathcal{S}_2 =  \langle \varGamma, \, \{ C_i \} _{i=1}^{n},$  $ \, \{ D_i \} _{i=1}^{m_2} \rangle $ with size $s_2$,   we have
		\begin{align} \label{Minkowski sum of spectrahedral shadows}
			\!	\mathcal{S}_1  \oplus \mathcal{S}_2 =  \langle  \tilde{\varLambda},\, \{ \tilde{A}_i \} _{i=1}^{n}, \, \{ \tilde{B}_i \}_{i=1}^{\tilde{m}} \rangle,
		\end{align}
		where $\tilde{m} = n+m_1+m_2$, $\tilde{\varLambda}, \tilde{A}_i, \tilde{B}_i \in \mathbb{S}^{s_1+s_2} $, $\tilde{\varLambda} = \mathrm{diag}(\varLambda , \,\varGamma ) $,  $\tilde{A}_i = \mathrm{diag}(\textbf{0} , \,  A_i) $  and
		\begin{equation} \notag
			\begin{aligned}
				\tilde{B}^i =  \begin{cases}
					\mathrm{diag}( \, -A_{i}, \, C_{\,i}),  \;\;\, 1\leq i \leq n\\
					\mathrm{diag}( \, B_{i-n}, \, \textbf{0}),  \;\;\, n+1\leq i \leq n+m_1\\
					\mathrm{diag}(\textbf{0} , \,  D_{i-n-m_1}),  \;\;\, n+m_1+1\leq i \leq \tilde{m}\\
				\end{cases} \hspace{-8pt}.
			\end{aligned}
		\end{equation} 
	The  computational complexity  is  $\mathcal{O}(ns_1^2)$. \vspace{3pt}
\end{Propos}

	\begin{Pro}
		It is observed that $z = x+u, \, x \in \mathcal{S}_1,\, u \in \mathcal{S}_2$ iff $\exists \, y\in \mathbb{R}^{m_1}, \, \varLambda +\sum_{i=1}^n{(z_i-u_i)A_i}+\sum_{i=1}^{m_1}{y_iB_i}\succeq 0 $ and  $\exists \, v\in \mathbb{R}^{m_2}, \, \varGamma +\sum_{i=1}^n{u_iC_i}+\sum_{i=1}^{m_2}{v_iD_i}\succeq 0 $. Let $e = [u,\, y, \, v]^T$. The above  is further equivalent to $\exists \, e\in \mathbb{R}^{\tilde{m}}, \, \tilde{\varLambda} +\sum_{i=1}^n{z_i\tilde{A}_i}+\sum_{i=1}^{\tilde{m}}{e_i\tilde{B}_i}\succeq 0 $ and hence yields \eqref{Minkowski sum of spectrahedral shadows}.\vspace{1pt}
		
		\textit{Complexity:} The complexity to compute $- A^i$  for all  $1\leq i \leq n$ is  $\mathcal{O}(ns_1^2)$. The remaining operation is matrix concatenation    and hence the overall complexity is   $\mathcal{O}(ns_1^2)$.
	\end{Pro}

\subsubsection{Intersection} In the literature, \cite{nemirovski2006} has given the expression of intersection for spectrahedral shadows. Here we introduce this method as the following Lemma \ref{lemma of intersection}.
		\begin{Lem}[Intersection, {\cite[p. 428]{nemirovski2006}}] \label{lemma of intersection}
			Given  $ \mathcal{S}_1 =  \langle \varLambda,$  $\, \{ A_i \} _{i=1}^{n},\, \{ B_i \} _{i=1}^{m_1} \rangle $ with size $s_1$ and  $ \mathcal{S}_2 =  \langle \varGamma, \, \{ C_i \} _{i=1}^{n},$  $ \, \{ D_i \} _{i=1}^{m_2} \rangle $ with size $s_2$,   we have
		\begin{align} \label{intersection of spectrahedral shadows}
			\!	\mathcal{S}_1  \cap \mathcal{S}_2 =  \langle  \tilde{\varLambda},\, \{ \tilde{A}_i \} _{i=1}^{n}, \, \{ \tilde{B}_i \}_{i=1}^{\tilde{m}} \rangle,
		\end{align}
		where $\tilde{m} = m_1+m_2$, $\tilde{\varLambda} = \mathrm{diag}(\varLambda , \,\varGamma ) \in \mathbb{S}^{s_1+s_2}$, $ \tilde{A}_i = \mathrm{diag}(A_i, \,  C_i)\in \mathbb{S}^{s_1+s_2} $ and 
			\begin{equation} \notag
			\begin{aligned}
				\tilde{B}^i =  \begin{cases}
					\mathrm{diag}(  B_{i}, \,  \textbf{0}),  \;\;\, 1\leq i \leq m_1\\
					\mathrm{diag}(\textbf{0}, \,   \, D_{i-m_1}),  \;\;\, m_1+1\leq i \leq \tilde{m}\\
				\end{cases} \hspace{-8pt}.
			\end{aligned}
		\end{equation} 
		The  computational complexity  is  $\mathcal{O}(1)$. \vspace{3pt}
	\end{Lem}	
	
	\begin{Pro}
		The expression of \eqref{intersection of spectrahedral shadows} 	directly follows from Definition \ref{def of spectrahedral shadows} and \cite[p. 428]{nemirovski2006}.\vspace{1pt}
		
		\textit{Complexity:}  Constructing $\tilde{\varLambda}$, $\tilde{A}_i $  and $	\tilde{B}_i $ only involves  matrix concatenation.
	\end{Pro}

\subsubsection{Cartesian Product} Then we introduce the Cartesian product for  spectrahedral shadows.
    	\begin{Propos}[Cartesian product] \label{propos of Cartesian product}
    	Given  $ \mathcal{S}_1 =  \langle \varLambda,$  $\, \{ A_i \} _{i=1}^{n_1},\, \{ B_i \} _{i=1}^{m_1} \rangle $ with size $s_1$ and  $ \mathcal{S}_2 =  \langle \varGamma, \, \{ C^i \} _{i=1}^{n_2},$  $ \, \{ D^i \} _{i=1}^{m_2} \rangle $ with size $s_2$,   we have
    	\begin{align} \label{Cartesian product of spectrahedral shadows}
    		\!	\mathcal{S}_1  \times \mathcal{S}_2 =  \langle  \tilde{\varLambda},\, \{ \tilde{A}^i \} _{i=1}^{\tilde{n}}, \, \{ \tilde{B}^i \}_{i=1}^{\tilde{m}} \rangle,
    	\end{align}
    	where $\tilde{n} = n_1+n_2$, $\tilde{m} = m_1+m_2$, $\tilde{\varLambda}, \tilde{A}^i, \tilde{B}^i \in \mathbb{S}^{s_1+s_2} $, $\tilde{\varLambda} = \mathrm{diag}(\varLambda , \,\varGamma) $ and 
    	\begin{equation} \notag
    		\begin{aligned}
    			\tilde{A}^i& =  \begin{cases}
    				\mathrm{diag}(A^{i}, \,  \textbf{0}),  \;\;\, 1\leq i \leq n_1\\
    				\mathrm{diag}(\textbf{0}, \,   \, C^{i-n_1}),  \;\;\, n_1+1\leq i \leq \tilde{n}\\
    			\end{cases} \\
    			\tilde{B}^i &=  \begin{cases}
    				\mathrm{diag}(  B^{i}, \,  \textbf{0}),  \;\;\, 1\leq i \leq m_1\\
    				\mathrm{diag}(\textbf{0}, \,   \, D^{i-m_1}),  \;\;\, m_1+1\leq i \leq \tilde{m}\\
    			\end{cases} \hspace{-8pt}.
    		\end{aligned}
    	\end{equation} 
    	The  computational complexity  is  $\mathcal{O}(1)$. \vspace{3pt}
    \end{Propos}
    
    \begin{Pro}
    	Assume that $x \in \mathcal{S}_1$, $u \in \mathcal{S}_2$ and $z=[x^T , u^T]^T$. Then it can be verified that  $x \in \mathcal{S}_1$ and $u \in \mathcal{S}_2$ iff $\exists \, e \in \mathbb{R}^{\tilde{m}},\, \tilde{\varLambda} +\sum_{i=1}^n{z_i\tilde{A}^i}+\sum_{i=1}^{\tilde{m}}{e_i\tilde{B}^i}\succeq 0 $, which yields \eqref{Cartesian product of spectrahedral shadows}. \vspace{1pt}
    	
    		\textit{Complexity:}  Constructing $\tilde{\varLambda}$, $\tilde{A}^i $  and $	\tilde{B}^i $ only involves  matrix concatenation.
    \end{Pro}

\subsubsection{Minkowski-Firey $\textit{L}_\textit{p}$ sum}
 	Minkowski-Firey $\textit{L}_\textit{p}$ sum is the concept first studied in  Brunn-Minkowski-Firey theory for the mixed volumes problem \cite{firey1962, lutwak1993, schneider2013}. Based on ellipsoids, recent work  \cite{halder2020}  introduced this set operation into RA. In the following part, we  prove that spectrahedral shadows  are closed under Minkowski-Firey $\textit{L}_\textit{p}$ sum. \vspace{2pt}

  \begin{Def} \label{def of p-sum}
 	Given $p \geq 1$ and $X, Y\subset \mathbb{R}^{n}$ with $\textbf{0} \in X \cap Y $,  the \textit{Minkowski-Firey $\textit{L}_\textit{p}$ sum} of $X$ and $Y$ is defined as
 	\begin{equation} \notag
 		\begin{aligned}
 		X +_p Y =  \{t^{{1}/{p^\prime}} x +(1-t)^{{1}/{p^\prime}}y\! : x \in X, \, y\in Y, \, 0\leq t \leq 1 \}
 		\end{aligned}
 	\end{equation} 
 	where $p^\prime$ is the  Hölder conjugate of $p$, i.e., $\frac{1}{p}+ \frac{1}{p^\prime} =1$.
 	\vspace{3pt}
 \end{Def}
 
 \begin{Rema}
   One  might notice that Definition \ref{def of p-sum} is different from the classical definition of Minkowski-Firey $\textit{L}_\textit{p}$ sum as defined by  support functions, e.g., \cite{firey1962, schneider2013}. Indeed, the classical definition is only valid for compact convex sets. It was shown in \cite{lutwak2012}  that Definition \ref{def of p-sum} generalizes the classical definition to the case of arbitrary subsets in Euclidean space. In this paper, we adopt  Definition \ref{def of p-sum} since spectrahedral shadows are not necessarily compact sets (see  Example \ref{example of noncompact}).	\vspace{3pt}
 \end{Rema}
 
\begin{figure}[!t]
 	\centering
 	\subfigure{
 		\hspace{-6pt} \includegraphics[scale=0.45]{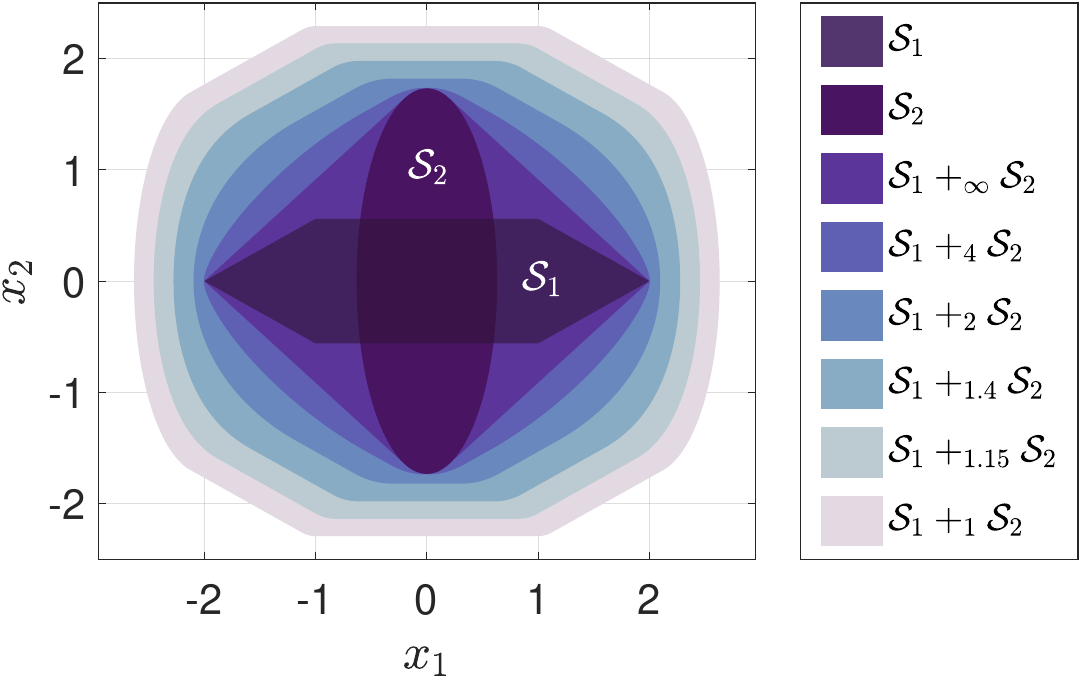}
 	}

 	\vspace{-4pt}
 	\caption{Minkowski-Firey $\textit{L}_\textit{p}$-sums of a zonotope $\mathcal{S}_1$ and an ellipsoid $\mathcal{S}_2$ for $p \in\{ 1, 1.15, 1.4, 2, 4, \infty\}$.} \label{fig4}
 \end{figure}

 For sets containing the origin, the Minkowski sum and convex hull  are two special cases of Minkowski-Firey $\textit{L}_\textit{p}$-sum when $p=1$ and $p=\infty$, respectively. Moreover, the  Minkowski-Firey $\textit{L}_\textit{p}$-sum of two  sets $X_1$ and $X_2$ has the following inclusion relationship $$
		X_1 +_1 X_2  \supseteq  	X_1 +_{p_1} X_2   \supseteq     	X_1 +_{p_2} X_2 \supseteq  	X_1 +_{\infty} X_2 $$
 for arbitrary $1\leq p_1 \leq p_2 \leq \infty$ \cite{firey1962}.   Fig \ref{fig4} gives an  example to illustrate  the  Minkowski-Firey $\textit{L}_\textit{p}$-sum of two sets with the increasing value of $p$. Then we introduce Lemma \ref{quasi q lemma} as the prerequisite  for the subsequent Proposition \ref{propos of p-sum}.
 
   \begin{Lem}[\cite{ben2001}] \label{quasi q lemma}
 	Given arbitrary rational $q$  with $ 0  \leq  q \leq 1$,  the set $\{x \in \mathbb{R}^2 \! :   x_1^q \geq x_2 \geq 0\}$ is exactly $\mathcal{S}(q) =$
 	\begin{equation} \notag
 			\resizebox{1.01\hsize}{!}{
 			$	
 			 \left\{x \in \mathbb{R}^2 \! : \!\! \begin{array}{ll}
 			\, 	t^{0}_{j}= \begin{cases}
 					x_1,  \;\; 1\leq j \leq c_1\\
 					 \, 1,  \;\;\; c_1+1\leq j \leq c_2\\
 					x_2 ,  \;\; c_2+1\leq j \leq 2^l\\
 				\end{cases} \hspace{-10pt}, \;\; t_1^l = x_2 , \\ \noalign{\vskip 5pt}
 				 \left[ \begin{matrix}
 					\,\frac{t^{i-1}_{2j-1}+t^{i-1}_{ 2j}}{2} \hspace{-8pt}      &	0 & t^i_{ j}  \vspace{2pt}  \,  \\
 					\,*  &	\frac{t^{i-1}_{2j-1}+t^{i-1}_{ 2j}}{2}  \hspace{-4pt} &   \frac{t^{i-1}_{2j-1}-t^{i-1}_{ 2j}}{2} \vspace{2pt} \,\\
 					\,* &	* & \frac{t^{i-1}_{2j-1}+t^{i-1}_{ 2j}}{2}   \,   
 				\end{matrix} \right] \! \succeq 0,  \begin{array}{ll} \forall \, 1\leq i \leq l \\  \forall \, 1\leq j \leq 2^{l-i} 
 					\end{array} \end{array}  \hspace{-10pt} \right\}
 		$}
 	\end{equation} 
 	where $c_1, c_2 \in \mathbb{N}$, $q = \frac{c_1}{c_2}$ is the fraction  in lowest terms, and $l$ is the smallest integer such that $2^l\geq c_2$. 
 	\vspace{3pt}
	 \end{Lem}
 
  	\begin{Pro}
 	By the construction technique  in \cite[Lecture 3.3.1]{ben2001}, $\{x \in \mathbb{R}^2 \! :   x_1^q \geq x_2 \geq 0\}$ is formulated as a conic quadratic-representable set. Then we  obtain Lemma \ref{quasi q lemma} using the well-known  method to convert general conic quadratic-representable sets to spectrahedral shadows \cite[(4.2.1)]{ben2001}.
 \end{Pro}

   	In  Lemma \ref{quasi q lemma},  all linear equations defining $\mathcal{S}(q)$  can be eliminated through variable substitution. Namely, $\mathcal{S}(q)$ is only defined by LMIs, which means that $\mathcal{S}(q)$  is  a spectrahedral shadow. Due to space limitations and for the sake of clarity,  we write $\mathcal{S}(q)$  in the form of Lemma \ref{quasi q lemma} instead of Definition \ref{def of spectrahedral shadows}. In practice, once the value of $p$ is determined,     $\mathcal{S}(q)$  can be  directly written as the form of Definition \ref{def of spectrahedral shadows} according to Lemma \ref{quasi q lemma}.  \vspace{3pt}	
   	
   	  \begin{Propos}[Minkowski-Firey $\textit{L}_\textit{p}$ sum] \label{propos of p-sum}
   		Given  $ \mathcal{S}_1 =  \langle \varLambda,$  $\, \{ A_i \} _{i=1}^{n},\, \{ B_i \} _{i=1}^{m_1} \rangle $ with size $s_1$,  $ \mathcal{S}_2 =  \langle \varGamma, \, \{ C_i \} _{i=1}^{n},$  $ \, \{ D_i \} _{i=1}^{m_2} \rangle $ with size $s_2$, $\textbf{0} \in \mathcal{S}_1 \cap \mathcal{S}_2$ and arbitrary rational $p$  with  $p \geq 1$,  we have
   			\begin{align} \label{p-sum of spectrahedral shadows}
   					\!	\mathcal{S}_1  +_p \mathcal{S}_2 =  \{ x&\in \mathbb{R}^n \!  : \, \exists \, y\in \mathbb{R}^m,\; [y_1, y_3]^T\! , [y_2, y_4]^T \!\! \in  \mathcal{S}(q) \nonumber \\& \tilde{\varLambda} +\sum_{i=1}^n{x_i\tilde{A}_i}+\sum_{i=1}^m{y_i\tilde{B}_i}\succeq 0  \}.
   			\end{align}
   		In \eqref{p-sum of spectrahedral shadows},  $m= n+m_1+m_2+4$, $q= 1-\frac{1}{p}$,  $\mathcal{S}(q)$ is the spectrahedral shadow constructed by applying Lemma \ref{quasi q lemma},  and
   		 \begin{flalign} 
   			&\ \tilde{\varLambda} = \mathrm{diag}(\left[ \begin{matrix}
   				\,1  \hspace{-4pt}   &	0  \,  \\
   				\,* \hspace{-4pt}  &	-1  \
   			\end{matrix} \right]\! , \, \textbf{0}),& \notag\\
   		&\	\tilde{A}_i = \mathrm{diag}(\textbf{0}_{2\times2} , \,  A_i, \, \textbf{0}) ,\; i =1,2,..,n,  & 			\notag
   	\end{flalign} 
   		\begin{equation} \notag
   		\begin{aligned}
   		 \tilde{B}_i =  \begin{cases}
   		\mathrm{diag}(\left[ \begin{matrix}
   			\,-1  \hspace{-4pt}   &	0  \,  \\
   			\,* \hspace{-4pt}  &	1  \
   		\end{matrix} \right] \! ,  \, \textbf{0}), \;\;\,  1\leq i \leq 2\\
   		 \mathrm{diag}(\textbf{0}_{2\times2}, \,   \varLambda^{+}, \, \textbf{0}),  \;\;\, i=3\\
   		 \mathrm{diag}(\textbf{0} ,  \, \varGamma^{+}) ,  \;\;\, i=4\\
   		 \mathrm{diag}(\textbf{0} , \, -A_{i-4}, \, C_{\,i-4}),  \;\;\, 5\leq i \leq n+4\\
   		 \mathrm{diag}(\textbf{0}_{2\times2} , \, B_{i-n-4}, \, \textbf{0}),  \;\;\, n+5\leq i \leq n+m_1+4\\
   		 \mathrm{diag}(\textbf{0} , \,  D_{i-n-m_1-4}),  \;\;\, n+m_1+5\leq i \leq m\\
   		\end{cases} \hspace{-8pt},
   			\end{aligned}
   		\end{equation} 
   		where $\tilde{\varLambda}, \tilde{A}_i ,\tilde{B}_i \in  \mathbb{S}^{2+s_1+s_2}$, and $\varLambda^{+}$ and  $\varGamma^{+}$ are obtained by choosing $y^* \in \mathbb{R} ^{m_1}$ and $v^* \in \mathbb{R}^{m_2}$  such that $\varLambda^{+}= \varLambda +\sum_{i=1}^{m_1}{y^*_iB_i}\succeq 0$ and $\varGamma^{+} = \varGamma +\sum_{i=1}^{m_1}{y^*_iD_i}\succeq 0$, respectively. The overall  computational complexity  is $\mathcal{O}(ns_1^2+\sqrt{s_1}(m_1s_1^2+m_1^\omega+s_1^\omega)+\sqrt{s_2}(m_2s_2^2+m_2^\omega+s_2^\omega))$.
   		\vspace{3pt}
   	\end{Propos}
   	
   	\begin{Pro}
   		Due to $\textbf{0} \in \mathcal{S}_1 $, \vspace{1pt} there always exists $y^*$ such that $\varLambda^{+} = \varLambda +\sum_{i=1}^{m_1}{y^*_iB_i}\succeq 0$, and 	\begin{equation} \notag
   			\begin{aligned}
   		\exists \, y\in \mathbb{R} ^m ,\; \varLambda +&\sum_{i=1}^n{x_iA_i} +\sum_{i=1}^m{y_iB_i}\succeq 0 \\   \xLeftrightarrow{ \;\, \tilde{y} = y -y^*  } \; &\exists \, \tilde{y} \in \mathbb{R} ^m ,\; \varLambda^{+} +\sum_{i=1}^n{x_iA_i}+\sum_{i=1}^m{\tilde{y}_iB_i}\succeq 0 ,
   			\end{aligned}
   		\end{equation} 
   		which implies \vspace{1pt} $\mathcal{S}_1 = \langle \varLambda^{+} , \, \{ A_i \}_{i=1}^{n},\, \{ B_i \} _{i=1}^{m_1} \rangle$. Similarly,\vspace{1pt} we have $\mathcal{S}_2 =\langle \varGamma^{+} , \, \{ C_i \} _{i=1}^{n},\, \{ D_i \} _{i=1}^{m_2} \rangle$. For any $r \in 	\mathcal{S}_1  +_p \mathcal{S}_2$, there exists $x\in \mathcal{S}_1$  and $u\in \mathcal{S}_2$ such that  
   		\begin{equation} \label{formula of p sum}
   			\begin{aligned}
   		r = t_1^{{1}/{p^\prime}}x + t_2^{{1}/{p^\prime}} u, \; \exists \, t_1, t_2 \geq 0, \, t_1+t_2 = 1.
   			\end{aligned}
   		\end{equation} 
   		Let $\bar{x} = t_1^{{1}/{p^\prime}}x$  and $\bar{u} = t_2^{{1}/{p^\prime}}u$. It follows from \eqref{formula of p sum} that $r \in \mathcal{S}_r$, where 
   		 \begin{equation} \notag
   		 	\resizebox{1.01\hsize}{!}{
   		 		$	
   		 		 \mathcal{S}_r=	\! \left\{ \bar{x} + \bar{u}  \!: \!\! \begin{array}{ll}
   		 					\exists \, \bar{y}\in \mathbb{R}^{m_1} ,\;  \bar{v}\in \mathbb{R}^{m_2} ,\; t_1, t_2 \geq 0, \, t_1+t_2 = 1 \\ t_1^{{1}/{p^\prime}}\varLambda^{+} +\sum_{i=1}^n{\bar{x}_iA_i} +\sum_{i=1}^{m_1}{\bar{y}_iB_i}\succeq 0, \\
   		 						   t_2^{{1}/{p^\prime}}\varGamma^{+} +\sum_{i=1}^n{\bar{u}_iC_i} +\sum_{i=1}^{m_2}{\bar{v}_iD_i}\succeq 0
   		 				 \end{array} \hspace{-5pt}  \right\}.
   		 	$ }
   		 	\end{equation}
   		 Thus, $\mathcal{S}_1  +_p \mathcal{S}_2 \subseteq \mathcal{S}_r$. Note that the  reverse  of the above derivation (i.e., $\mathcal{S}_r \subseteq  \mathcal{S}_1  +_p \mathcal{S}_2 $)  holds true as long as $t_1, t_2 \neq 0$. Consider the remaining case of $t_1 = 0$ (and hence $t_2 = 1$) without loss of generality.  For arbitrary  $\bar{r} \in \mathcal{S}_r$, there exists $\bar{x}, \bar{y}, \bar{u}, \bar{v}$ satisfying $\sum_{i=1}^n{\bar{x}_iA_i}+\sum_{i=1}^{m_1}{\bar{y}_iB_i}\succeq 0$ and $\varGamma^{+} +\sum_{i=1}^n{\bar{u}_iC_i} +\sum_{i=1}^{m_2}{\bar{v}_iD_i}\succeq 0$ such that $	\bar{r}=  \bar{x} + \bar{u} $. Due to $\bar{x} \in \mathcal{S}_1$ (note $\varLambda^{+} \succeq 0$),  $\bar{u} \in \mathcal{S}_2$ and $t_2 =1$, the condition \eqref{formula of p sum} holds for $\bar{r}$, i.e.,  $\bar{r} \in \mathcal{S}_1  +_p \mathcal{S}_2$. Hence, $\mathcal{S}_r \subseteq  \mathcal{S}_1  +_p \mathcal{S}_2 $ also holds for $t_1 = 0$ or $t_2 = 0$. It follows from $\mathcal{S}_1  +_p \mathcal{S}_2 \subseteq \mathcal{S}_r$ that $\mathcal{S}_r  = \mathcal{S}_1  +_p \mathcal{S}_2$. Due to $\varLambda^{+}, \varGamma^{+} \succeq 0$, $\mathcal{S}_r $ is equivalent to
   		   		 \begin{equation}\label{formula of p sum 2}
   		 	\resizebox{0.88\hsize}{!}{
   		 		$	
   		 			\! \left\{ \bar{x} + \bar{u}  \!: \!\! \begin{array}{ll}
   		 			\exists \, \bar{y}\in \mathbb{R}^{m_1} ,\;  \bar{v}\in \mathbb{R}^{m_2} , \, t_1+t_2 = 1 \\ t_3\varLambda^{+} +\sum_{i=1}^n{\bar{x}_iA_i} +\sum_{i=1}^{m_1}{\bar{y}_iB_i}\succeq 0, \\
   		 		t_4\varGamma^{+} +\sum_{i=1}^n{\bar{u}_iC_i} +\sum_{i=1}^{m_2}{\bar{v}_iD_i}\succeq 0, \\
   		 			t_1^{{1}/{p^\prime}} \geq t_3 \geq 0 ,\;  	t_2^{{1}/{p^\prime}} \geq t_4 \geq 0
   		 		\end{array} \hspace{-5pt}  \right\}.
   		 		$ }
   		 \end{equation}
   		 Then \eqref{formula of p sum 2} is written as  \eqref{p-sum of spectrahedral shadows}  by analogy   with Proposition \ref{propos of Minkowski sum}.
   		 
   		 \textit{Complexity:}  Constructing $\mathcal{S}(q) $ as a spectrahedral shadow by Lemma \ref{quasi q lemma} only involves  matrix element assignment, where the size of the matrix is constant once the value of $q$ is determined.   Hence, the complexity is $\mathcal{O}(c(q))$, where $c(q)$ is a constant about $q$. Moreover, $\varLambda^{+}$ and  $\varGamma^{+}$ can be obtained by finding a feasible solution to an SDP problem, and the complexity is  $\mathcal{O}(\sqrt{s_1}(m_1s_1^2+m_1^\omega+s_1^\omega)+\sqrt{s_2}(m_2s_2^2+m_2^\omega+s_2^\omega))$. The complexity to compute $-A_i$ for all $1\leq i \leq n$ is  $\mathcal{O}(ns_1^2)$. The remaining operation is  matrix concatenation.  Omitting constant  and lower-order terms, the overall complexity is $\mathcal{O}(ns_1^2+\sqrt{s_1}(m_1s_1^2+m_1^\omega+s_1^\omega)+\sqrt{s_2}(m_2s_2^2+m_2^\omega+s_2^\omega))$.
   	\end{Pro}
   	
   	\begin{Rema}
   	 We can extend Proposition \ref{propos of p-sum}   to compute the Minkowski-Firey $\textit{L}_\textit{p}$ sums of $l$  spectrahedral shadows within $\mathcal{O}(ns_1^2+\sum_{i=1}^{l}\sqrt{s_i}(m_is_i^2+m_i^\omega+s_i^\omega))$ time, where $s_i$ and $m_i$ are the size and lifted dimension of the $i$-th added spectrahedral shadow. In the literature, \cite{halder2020} proposes an outer approximated method to compute  the Minkowski-Firey $\textit{L}_\textit{p}$ sum of $l$ $n$-dimensional ellipsoids within $\mathcal{O}(ln^3)$ time \cite[Section \uppercase\expandafter{\romannumeral6}-D]{halder2020}. We will see in  Section \ref{section: relation with existing set representation}  that $n$-dimensional ellipsoids are a special case of $n$-dimensional spectrahedral shadows with  size $n+1$ and lifted dimension $0$. This means that the task in \cite{halder2020} can be solved through Proposition \ref{propos of p-sum} within $\mathcal{O}(n(n+1)^2+\sum_{i=1}^{l}\sqrt{n+1} (n+1)^{\omega})$ = $\mathcal{O}(n^3+ln^{\omega+0.5})$ runtime without approximation. \vspace{3pt}
   	\end{Rema}

\subsubsection{Conic hull}
	 Then we introduce the \textit{conic hull}, which is an   operation commonly used  in convex analysis and optimization.\vspace{3pt}

	 \begin{Def} \label{def of conic hull}
		Given  $X \subset \mathbb{R}^{n}$,  the \textit{conic hull} of $X$  is defined as $	\mathrm{cone}(X) =  \{ t x : x \in X, \, t \geq 0 \} $.
		\vspace{3pt}
	\end{Def}
	
\begin{figure}[!t]
		\centering
		\subfigure[Conic hull]{
			\hspace{-7pt} \includegraphics[scale=0.235]{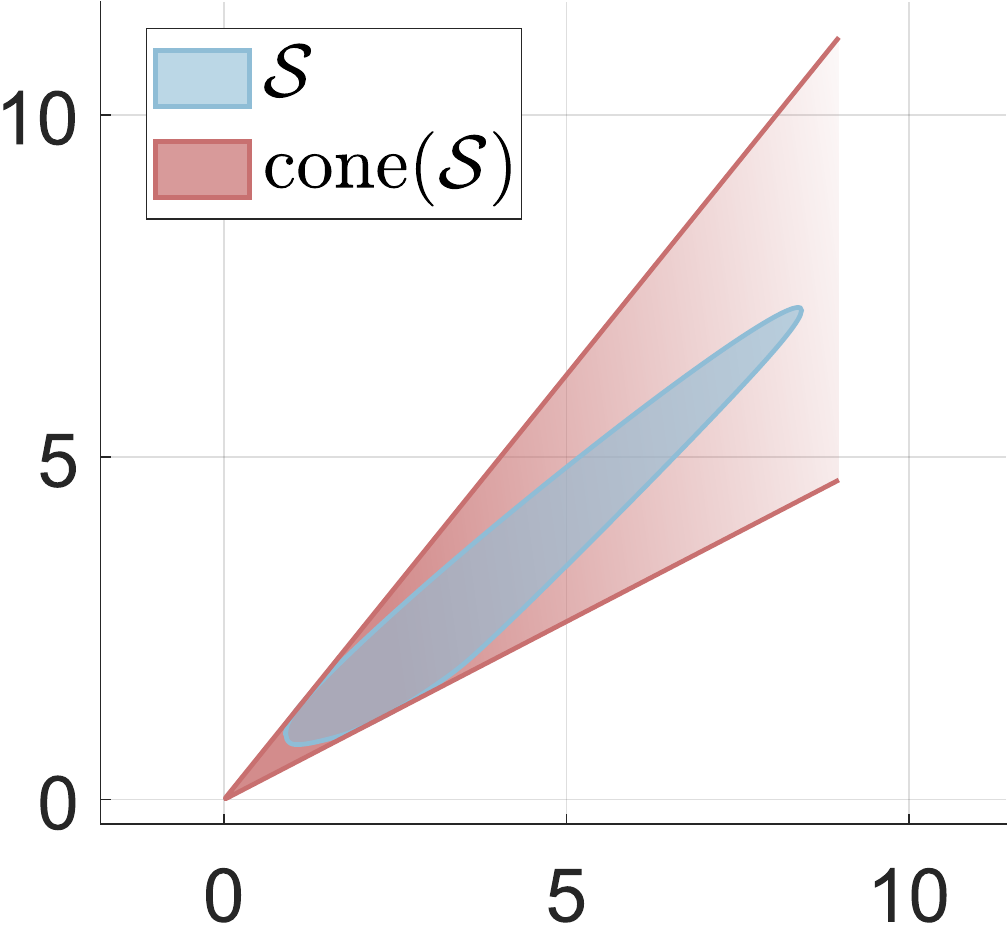}\label{fig5a}
		}
		\subfigure[Convex hull]{
			\hspace{-7pt} \includegraphics[scale=0.224]{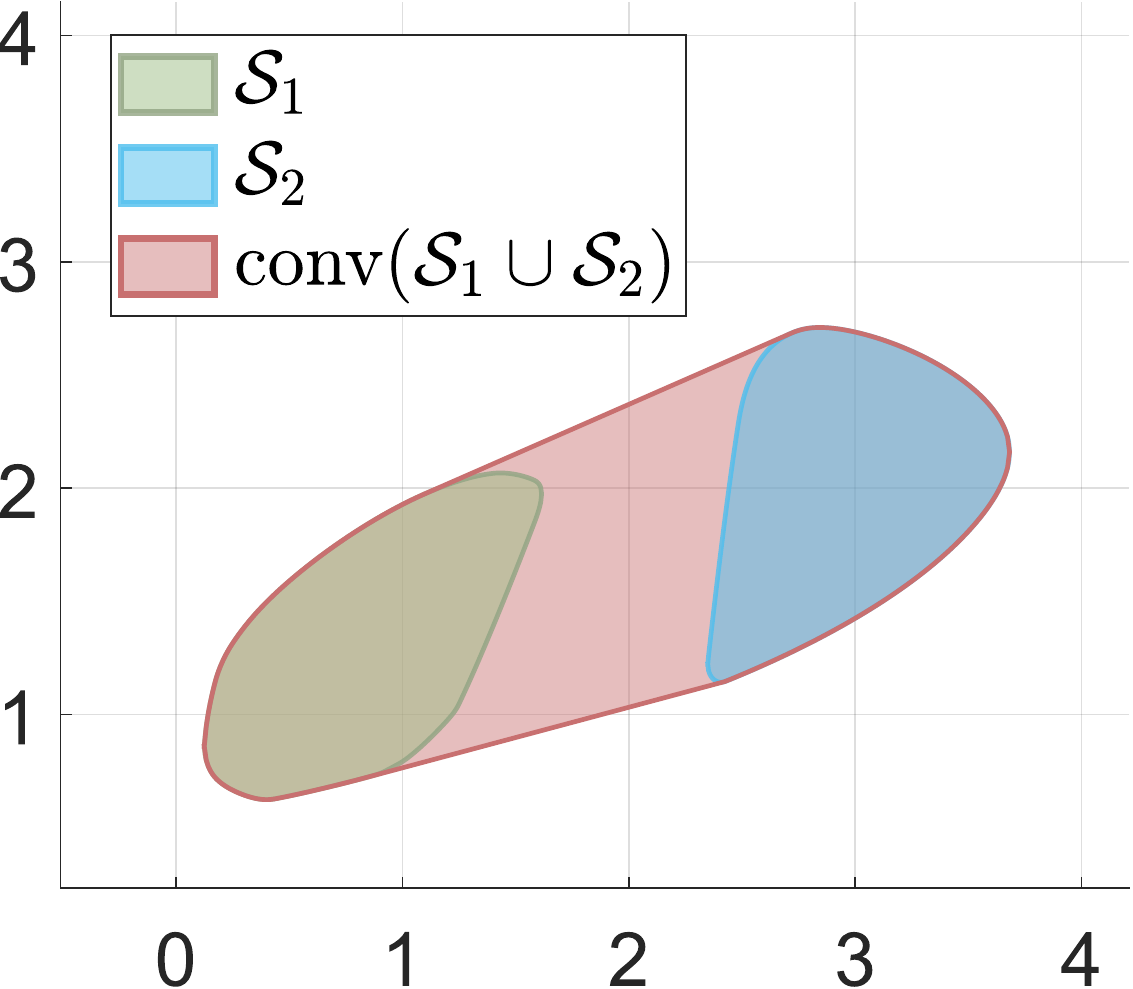}\label{fig5b}
		}
		\vspace{-4pt}
		\caption{The conic hull and convex hull of spectrahedral shadows.} \label{fig5}
	\end{figure}

	Geometrically, the conic hull of a set $X$ is the smallest cone that contains $X$. Fig. \ref{fig5a} shows an example of the conic hull of a spectrahedral shadows. Then Proposition \ref{propos of conic hull} shows how to compute the conic hull of spectrahedral shadows.
	
	\begin{Propos}[Conic hull] \label{propos of conic hull}
	For any nonempty $ \mathcal{S} =  \langle \varLambda,\,$ $ \{ A_i \} _{i=1}^{n}, \, \{ B_i \} _{i=1}^{m} \rangle $ with size $s$, there is a spectrahedral shadow $ \mathcal{S}_c =  \langle \textbf{0},\,\{ A_i^c \} _{i=1}^{n}, \, \{ B_i^c \} _{i=1}^{m+1} \rangle $ with 
	\begin{equation} \label{param of conic hull}
	\begin{aligned}
	&	  A_i^c = \mathrm{diag}(A_i , \, 0) \in \mathbb{S}^{s+1} ,\, \forall\, 1\leq i\leq n, \\
			& B_i^c =  \begin{cases}
				\mathrm{diag}(B_i \! ,  \, \textbf{0}), \;\;  1\leq i \leq m\\
				\mathrm{diag}( {\varLambda}, \, 1),  \;\; i=m+1\\
			\end{cases} \hspace{-8pt} \in \mathbb{S}^{s+1}
		\end{aligned}
	\end{equation} 
	such that
	\begin{equation} \label{boundary diff of conic hull}
		\begin{aligned}
		 \mathrm{cone}(\mathcal{S}) \subseteq \mathcal{S}_c, \;\,  \overline{\mathrm{cone}(\mathcal{S})} = \overline{\mathcal{S}_c}.
		\end{aligned}
	\end{equation} 
	Particularly, if $\textbf{0} \in   \mathcal{S}$ or $\mathcal{S}$ is a bounded set, then 	
	\begin{equation} \label{formula of conic hull}
		\begin{aligned}
		\mathrm{cone}(\mathcal{S}) = \mathcal{S}_c.
		\end{aligned}
	\end{equation} 
	The  computational complexity to obtain $ \mathcal{S}_c$  is  $\mathcal{O}(1)$. \vspace{3pt}
	\end{Propos}
	
	\begin{Pro}
    $\mathcal{S}_c$	can be rewritten as
    \begin{equation} \notag
    	\begin{aligned}
    	\{ x\in \mathbb{R}^n \!  : \, \exists \, y\in \mathbb{R}^m,\; t\geq0 ,\; {t}\varLambda +\sum_{i=1}^n{x_i{A}_i}+\sum_{i=1}^m{y_i{B}_i}\succeq 0  \}.
    	\end{aligned}
    \end{equation} 
    Similar to the proof of $\mathcal{S}_1  +_p \mathcal{S}_2 \subseteq \mathcal{S}_r$ in Proposition \ref{propos of p-sum}, we obtain $ \mathrm{cone}(\mathcal{S}) \subseteq \mathcal{S}_c$.  Then we  prove $ \overline{\mathrm{cone}(\mathcal{S})} = \overline{\mathcal{S}_c}$. Since $ \mathrm{cone}(\mathcal{S}) \subseteq \mathcal{S}_c$, we only need to prove ${\mathcal{S}_c} \subseteq \overline{\mathrm{cone}(\mathcal{S})} $.  For any $r \in \mathcal{S}_c $, there exist $y\in \mathbb{R}^m$ and $t\geq 0$ such that ${t}\varLambda +\sum_{i=1}^n{r_i{A}_i}+\sum_{i=1}^m{y_i{B}_i}\succeq 0 $. Let $r^{\epsilon} = r+ \epsilon r $, $t^{\epsilon} = t+ \epsilon t $ and $y^{\epsilon} = y+ \epsilon y $  with $\epsilon >0$. It is clear that $t^{\epsilon} > 0$  and ${t}^{\epsilon}\varLambda +\sum_{i=1}^n{r_i^{\epsilon}{A}_i}+\sum_{i=1}^m{y_i^{\epsilon}{B}_i}\succeq 0 $. Then there exist $x^{\epsilon}=\frac{r^{\epsilon}}{{t}^{\epsilon}} $ and $\tilde{y}^{\epsilon}=\frac{y^{\epsilon}}{{t}^{\epsilon}} $ satisfying $\varLambda +\sum_{i=1}^n{x_i^{\epsilon}{A}_i}+\sum_{i=1}^m{\tilde{y}_i^{\epsilon}{B}_i}\succeq 0 $, which implies $x^{\epsilon} \in \mathcal{S}$ and hence we have $r^{\epsilon}\in \mathrm{cone}(\mathcal{S}) $ according to Definition \ref{def of conic hull}. When $\epsilon \rightarrow +0$, we have  $r_\epsilon \rightarrow r$, and $r \in \overline{\mathrm{cone}(\mathcal{S})}$ holds due to the closure property of $\overline{\mathrm{cone}(\mathcal{S})} $. Thus, ${\mathcal{S}_c} \subseteq \overline{\mathrm{cone}(\mathcal{S})} $.
    
    Consider the case of $\textbf{0} \in   \mathcal{S}$. By analogy  with the proof of $\mathcal{S}_r \subseteq  \mathcal{S}_1  +_p \mathcal{S}_2 $ in  Proposition \ref{propos of p-sum}, we have $\mathcal{S}_c \subseteq \mathrm{cone}(\mathcal{S}) $, thereby leading to \eqref{formula of conic hull}. Then consider  the case where $\mathcal{S}$ is bounded. For arbitrary $r\in \mathcal{S}_c$, there exist $y\in \mathbb{R}^m$ and $t\geq 0$ such that ${t}\varLambda +\sum_{i=1}^n{r_i{A}_i}+\sum_{i=1}^m{y_i{B}_i}\succeq 0 $. When $t>0$, we  have $r\in \mathrm{cone}(\mathcal{S})$  by variable substitution. To prove $r\in \mathrm{cone}(\mathcal{S})$ for $t=0$, let $x^r= x^*+\alpha r$ with $\alpha \geq 0$ and $ x^*\in \mathcal{S}$. It follows from $\sum_{i=1}^n{r_i{A}_i}+\sum_{i=1}^m{y_i{B}_i}\succeq 0 $ and $\exists\, y^* \in \mathbb{R}^{m},\, \varLambda +\sum_{i=1}^n{x_i^*{A}_i}+\sum_{i=1}^m{y_i^*{B}_i}\succeq 0$ that   $\exists\, y^r = y^*+\alpha y,\, \varLambda +\sum_{i=1}^n{x_i^r{A}_i}+\sum_{i=1}^m{y^r_i{B}_i}\succeq 0$, i.e., $x^r \in \mathcal{S}$. If $r \neq \textbf{0}$, it is contradictory to state that $S$ is bounded, given that $$\lim_{\alpha \rightarrow +\infty} x^*+\alpha r = \lim_{\alpha \rightarrow +\infty} x^r \in \mathcal{S}. $$ Thus,  $r = \textbf{0} \in \mathrm{cone}(\mathcal{S})$ and we finally obtain  $\mathcal{S}_c \subseteq \mathrm{cone}(\mathcal{S}) $.\vspace{1pt}
    
     \textit{Complexity:} Constructing \eqref{param of conic hull} only needs   matrix concatenation. 
	\end{Pro}
	
	\begin{Rema} \label{remark of conic hull}
		A special case of Proposition \ref{propos of conic hull}, i.e., computing the closed conic hull of $\langle \varLambda,\,$ $ \{ A_i \} _{i=1}^{n}, \, \{ B_i \} _{i=1}^{m} \rangle$ under the implication  condition  $\sum_{i=1}^m{y_i{B}_i}\succeq 0 \Rightarrow y=\textbf{0}$, is  given in \cite[p. 428]{nemirovski2006}. Moreover,  from the perspective of convex analysis,  $\mathcal{S}_c$ given in  Proposition \ref{propos of conic hull}  is not  the exact conic hull of $\mathcal{S}$ when the condition  for \eqref{formula of conic hull} is invalid. But in practice, $\mathcal{S}_c$  can  be seen as the lossless conic hull of $\mathcal{S}$ even if the condition  for \eqref{formula of conic hull} is invalid, because $\mathcal{S}_c$ and $\mathrm{cone}(\mathcal{S})$  have the same interior and differ at most on the topological boundaries (see \eqref{boundary diff of conic hull}).  The same is true for the following  convex hull and polytopic map computation. \vspace{3pt}
	\end{Rema}
	
\subsubsection{Convex hull}
	Besides the conic hull, the \textit{convex hull} is also  commonly seen in convex analysis and optimization. 
	The problem of computing  the convex hull of  spectrahedral shadows has been addressed in \cite{helton2009}. Here we introduce the expression   as the following Lemma \ref{lemma of convex hull}.
	
	\begin{Lem}[Convex hull, {\cite[Theorem 2.2]{helton2009}}] \label{lemma of convex hull}
		For arbitrary nonempty $ \mathcal{S}_1 =  \langle \varLambda^1,\, \{ A_i \} _{i=1}^{n}, \, \{ B_i \} _{i=1}^{m_1} \rangle $ with size $s_1$ and $ \mathcal{S}_2 =  \langle \varLambda^2,\, \{ C_i \} _{i=1}^{n}, \, \{ D_i \} _{i=1}^{m_2} \rangle $ with size $s_2$, there is a spectrahedral shadow $ \mathcal{S}_v =  \langle {\varLambda}^v ,\,\{ A_i^v \} _{i=1}^{n}, \, \{ B_i^v \} _{i=1}^{m} \rangle $ with 
		  \begin{flalign} 
			&\ {\varLambda}^v = \mathrm{diag}(\left[ \begin{matrix}
				\,1  \hspace{-4pt}   &	0  \,  \\
				\,* \hspace{-4pt}  &	0  \,
			\end{matrix} \right]\! , \, \textbf{0} , \, \varLambda^2 ),& \notag\\
			&\	{A}_i^v = \mathrm{diag}(\textbf{0} , \, C_i),\; i =1,2,..,n,  & 			\notag
		\end{flalign} 
		\vspace{-10pt}
		\begin{equation} \notag
			\begin{aligned}
				 {B}_i^v =  \begin{cases}
					\mathrm{diag}(\left[ \begin{matrix}
						\,-1  \hspace{-4pt}   &	0  \,  \\
						\,* \hspace{-4pt}  &	1  \,
					\end{matrix} \right]\!,  \varLambda^1  ,  \, -\varLambda^2), \;\;\, \vspace{1pt}  i = 1\\
					\mathrm{diag}(\textbf{0} ,  \, A_{i-1}, \, -C_{\,i-1}),  \;\;\, 2\leq i \leq n+1\\
					\mathrm{diag}(\textbf{0}_{2\times2} ,  \, B_{i-n-1}, \, \textbf{0}),  \;\;\, n+2\leq i \leq n+m_1+1\\
					\mathrm{diag}(\textbf{0} ,  \,  D_{i-n-m_1-1}),  \;\;\, n+m_1+2\leq i \leq m\\
				\end{cases} \hspace{-10pt},
			\end{aligned}
		\end{equation} 
		such that
		\begin{equation} 
			\begin{aligned}
					\mathrm{conv}(\mathcal{S}_1\cup \mathcal{S}_2) \subseteq \mathcal{S}_v, \;\,  \overline{	\mathrm{conv}(\mathcal{S}_1\cup \mathcal{S}_2)} = \overline{\mathcal{S}_v},
			\end{aligned}
		\end{equation} 
		where $ {\varLambda}^v, {A}_i^v , {B}_i^v \in \mathbb{S}^{2+s_1+s_2}$ and $m =  n+m_1+m_2+1$. Particularly, if  $\mathcal{S}_1$ and $\mathcal{S}_2$  are both  bounded sets, then 	
		\begin{equation} \label{formula of convex hull}
			\begin{aligned}
					\mathrm{conv}(\mathcal{S}_1\cup \mathcal{S}_2) = \mathcal{S}_v.
			\end{aligned}
		\end{equation} 
		The  computational complexity to obtain $ \mathcal{S}_v$  is  $\mathcal{O}(ns_2^2)$. \vspace{3pt}
	\end{Lem}
	
	\begin{Pro}
		The expression of $ \mathcal{S}_v$  follows from  \cite[Theorem 2.2]{helton2009} and Proposition \ref{propos of Minkowski sum}. 
		
		\textit{Complexity:} Computing $-\varLambda^2$  and $-C_i$  for all $1\leq i \leq n$  has complexity  $\mathcal{O}(s_2^2)$ and $\mathcal{O}(ns_2^2)$, respectively. The  remaining operation is  matrix concatenation, and hence the overall complexity is  $\mathcal{O}(ns_2^2)$.
	\end{Pro}
	
	\subsubsection{Linear set-valued map} 

	 To propagate uncertainties on  uncertain LPV systems,  spectrahedral shadows should be able to represent the sets under \textit{linear set-valued maps}.   Intuitively, a  set-valued map    maps a point into a set. We give a formal statement of the linear set-valued map  in  Definition  \ref{def of linear set-valued map}. \vspace{2pt}
	
	\begin{Def}[Linear set-valued map] \label{def of linear set-valued map}
		 A set-valued function $\textit{\textbf{T}}: \mathbb{R}^n \rightarrow \mathbb{P}(\mathbb{R}^n) $ is called a \textit{linear set-valued map} if $\textit{\textbf{T}}(x+y) = \textit{\textbf{T}}(x) \oplus \textit{\textbf{T}}(y)$ and $\textit{\textbf{T}}(cx) = c\textit{\textbf{T}}(x)$ for all $x,y \in \mathbb{R}^n$ and $c\in \mathbb{R}$. A set $X \subset V$ under a linear set-valued map $ \textit{\textbf{T}}$ is defined as $\textit{\textbf{T}}(X) = \cup_{x\in X}\, \textit{\textbf{T}}(x)$.\vspace{3pt}
	\end{Def}
	
 The linear set-valued map can be further classified based	on the type of the set $\textit{\textbf{T}}(x)$, e.g., interval map \cite{alamo2005, althoff2010} ($\textit{\textbf{T}}(x)$ is an interval)  and polytopic map ($\textit{\textbf{T}}(x)$ is a polytope). Next, we derive the  expression  of spectrahedral shadows under the polytopic map, and the  polytopic map is the  common  form of linear set-valued map  on uncertain LPV systems (e.g., \cite{xu2019b, silvestre2022}).  \vspace{3pt}

	\begin{Propos}[Polytopic map] \label{propos of polytopic map}
	For $ T_1, \,T_2, \, ... \, ,\, T_h \in \mathbb{R}^{l \times n} $, $ \textit{\textbf{T}} = \mathrm{conv}\{T_1, \,T_2, \, ... \, ,\, T_h  \} \subset \mathbb{R}^{l \times n} $   and $ \mathcal{S} =  \langle \varLambda,\,$ $ \{ A_i \} _{i=1}^{n}, \, \{ B_i \} _{i=1}^{m} \rangle $ with size $s$, there is a spectrahedral shadows $\mathcal{S}_p =  \langle {\varLambda}^p ,\,\{ A_i^p \} _{i=1}^{l}, \, \{ B_i^p \} _{i=1}^{m_p} \rangle $ with 
	\begin{flalign} 
		&\ {\varLambda}^p = \mathrm{diag}(\textbf{0},\, \left[ \begin{matrix}
			\,1  \hspace{-4pt}   &	0  \,  \\
			\,* \hspace{-4pt}  &	-1  \,
		\end{matrix} \right]\!   ),& \notag\\
		&\	{A}_i^p = \mathrm{diag}({A}_i^1, \, \textbf{0} ),\; i =1,2,..,l, \notag\\
		 & \,\{ B_i^p \} _{i=1}^{m_p} = \{M_1,...,M_h, N^2_1,...,N^2_l,N^3_1,...,N^3_l,...,N^h_l, \notag \\
		 & \hspace{57pt}  O^1_1,...,O^1_{m_1},O^2_1,...,O^2_{m_2},...,O^h_{m_h} \} 			\notag
	\end{flalign} 
	such that
	\begin{equation}   \label{boundary diff of polytopic map}
		\begin{aligned}
			\textit{\textbf{T}}(\mathcal{S})\subseteq \mathcal{S}_p, \;\,  	 \overline{\textit{\textbf{T}}(\mathcal{S})} = \overline{\mathcal{S}_p},
		\end{aligned}
	\end{equation} 
	where  $ {\varLambda}^p, {A}_i^p , {B}_i^p \in \mathbb{S}^{s_p}$, $s_p = 2+h+\sum\nolimits_{i=1}^h{s_i}$, $m_p = h+ l(h-1)+\sum\nolimits_{i=1}^h{m_i}$, $s_i$ is the corresponding size \vspace{1pt} of $\langle \varLambda^i,\, \{ A^i_j \} _{j=1}^{l}, \, \{ B^i_j \} _{j=1}^{m_i} \rangle$ ($1 \leq i \leq h$), and 
	\begin{equation} \notag
		\begin{aligned}
			&\langle \varLambda^i,\, \{ A^i_j \} _{j=1}^{l}, \, \{ B^i_j \} _{j=1}^{m_i} \rangle = T_i \circ \mathcal{S},\; \forall  \, 1\leq i \leq h, \\
			& M_i=\mathrm{diag}\left(\,  \underset{i-1}{\underbrace{0,...,0}},1,\hspace{-10pt} \underset{h-i+\sum\nolimits_{k=1}^h{\! s_k}}{\underbrace{0,...,0}}\hspace{-10pt},-1,1 \, \right)\!,\; \forall  \, 1\le i\le h , \\
			& N^i_j=\mathrm{diag}\left( -A^{1}_{j},\underset{\sum\nolimits_{k=2}^{i-1}{s_k}}{\underbrace{0,...,0}},A^{i}_{j},\hspace{-14pt}\underset{2+h+\sum_{k=i+1}^h{s_k}}{\underbrace{0,...,0}} \right)\!, \! \begin{array}{ll} \forall \, 2\leq i \leq h \\  \forall \, 1\leq j \leq l 
			\end{array}\!\!,\\
			& O^i_j=\mathrm{diag}\left(\underset{\sum\nolimits_{k=1}^{i-1}{s_k}}{\underbrace{0,...,0}},B^{i}_{j},\hspace{-14pt}\underset{2+h+\sum\nolimits_{k=i+1}^h{s_k}}{\underbrace{0,...,0}} \right)\!, \! \begin{array}{ll} \forall \, 1\leq i \leq h \\  \forall \, 1\leq j \leq m_i
			\end{array}\!\!.
		\end{aligned}
	\end{equation} 
	 Particularly, if $\textbf{0} \in  \mathcal{S}$ or $S$ is a bounded set, then
		\begin{equation} \label{formula of polytopic map}
		\begin{aligned}
		\textit{\textbf{T}}(\mathcal{S}) = \mathcal{S}_p.
		\end{aligned}
	\end{equation} 
	The complexity to obtain $ \mathcal{S}_p$  is  $\mathcal{O}(h\tilde{l}^4+ h\tilde{l}^3s + h\tilde{l}^2s^2 )$, where $\tilde{l} = \max\{l, n\}$ \vspace{3pt}
\end{Propos}

 \begin{Pro}
 	We first prove that $\textit{\textbf{T}}(\mathcal{S}) = \mathrm{conv}( \cup_{i=1}^{h}  \; T_i \circ \mathcal{S} )$ by noticing $ \textit{\textbf{T}} = \mathrm{conv}\{T_1, \,T_2, \, ... \, ,\, T_h  \}  $ and
 	\begin{equation} \notag
       	\begin{aligned}
       	z\in \textit{\textbf{T}}(\mathcal{S}) \,& \Leftrightarrow  \, z = \textit{\textbf{T}}(x), \, \exists\,  x \in \mathcal{S} \Leftrightarrow   z = Tx, \, \exists \,  T\in \textit{\textbf{T}}, \, x \in \mathcal{S} \\
       	& \Leftrightarrow  \, z = \sum_{i=1}^{h} \theta_i (T_ix), \,  \exists \,  \theta_i \geq 0,\, \sum_{i=1}^{h} \theta_i =1, \, x \in \mathcal{S} \\
       	    	& \Leftrightarrow  \, 	z\in  \mathrm{conv}( \cup_{i=1}^{h}  \; T_i \circ \mathcal{S} ).
       \end{aligned}
 	\end{equation}
 	Denote $ \mathcal{S}_i^T = T_i \circ \mathcal{S} = \langle \varLambda^i,\, \{ A^i_j \} _{j=1}^{l}, \, \{ B^i_j \} _{j=1}^{m} \rangle $.  Then $\mathcal{S}_i^T$ can be computed by Proposition \ref{propos of linear  map}. According to the convex hull expression given in \cite[Theorem 2.2]{helton2009}, there is a set
 		\begin{equation} \notag
 		\begin{aligned}
 			S_p=\left\{ \sum_{i=1}^h{u^i}\!:\hspace{-5pt} \begin{array}{c}
 				\exists \, v^i \in \mathbb{R}^{m_i}, \,	 t \in \mathbb{R}^{h}, \,t_i\ge 0,\sum_{i=1}^h{t_i}=1\\
 			\begin{array}{c}
 				t_1\Lambda ^1+\sum_{i=1}^l{u_{i}^{1}A_{i}^{1}}+\sum_{i=1}^{m_1}{v_{i}^{1}B_{1}^{i}}\succeq 0,  \\\noalign{\vskip 2pt}
 				t_2\Lambda ^2+\sum_{i=1}^l{u_{i}^{2}A_{i}^{2}}+\sum_{i=1}^{m_2}{v_{i}^{2}B_{2}^{i}}\succeq 0,\\
 				\vdots\\
 				t_h\Lambda ^h+\sum_{i=1}^l{u_{i}^{h}A_{i}^{h}}+\sum_{i=1}^{m_h}{v_{i}^{h}B_{h}^{i}}\succeq 0\\
 			\end{array}\\
 		\end{array} \! \right\}.
 		\end{aligned}
 	\end{equation}
 	such that  $\textit{\textbf{T}}(\mathcal{S}) = \mathrm{conv}( \cup_{i=1}^{h}  \; \mathcal{S}_i^T ) \subseteq \mathcal{S}_p$, $\overline{\textit{\textbf{T}}(\mathcal{S})} = \overline{\mathcal{S}_p}$, and $\textit{\textbf{T}}(\mathcal{S}) = \mathcal{S}_p$ when $\mathcal{S}_i^T$ for  $1\leq i \leq h$ are all bounded. Due to $ \mathcal{S}_i^T = T_i \circ \mathcal{S}$,  the boundedness condition for all $\mathcal{S}_i^T$  can be simplified as  the boundedness  for  $\mathcal{S}$. For the case of $\textbf{0} \in  \mathcal{S}$, we have $\textbf{0} \in \mathcal{S}_i^T$, then we can also prove $\textit{\textbf{T}}(\mathcal{S}) = \mathcal{S}_p$   by analogy  with the proof of Proposition \ref{propos of p-sum}. \vspace{1pt} Let $x = \sum_{i=1}^h{u^i} \in \mathbb{R}^{l}$.  Then $S_p$ is equivalently written as
 	 		\begin{equation} \notag
 	\resizebox{1.01\hsize}{!}{
 		$	
 		\hspace{-2pt}	S_p=\left\{ x\!: \hspace{-5pt}
 			\begin{array}{c}
 				\exists \, t \in \mathbb{R}^{h}, \,	 u^2\!, \, u^3\!,..., u^h \in \mathbb{R}^{l},  v^i \in \mathbb{R}^{m_i}, \, 1\leq i \leq h \\
 				\begin{array}{c}
 				{\varLambda}^p+\sum_{i=1}^n{x_iA_i^p}+\sum_{i=1}^h{t_iM_i} \\\noalign{\vskip 2pt}
 				+ \sum_{i=2}^h{\sum_{j=1}^l{u_j^iN_j^i}} + \sum_{i=1}^h{\sum_{j=1}^{m_i}{v_j^iO_j^i}} \succeq 0	\\
 				\end{array}\\
 			\end{array} \hspace{-5pt}\right\},
 		$}
 	\end{equation}
 	which yields Proposition \ref{propos of polytopic map}.\vspace{1pt}
 	
 	\textit{Complexity:} According to Proposition \ref{propos of linear  map}, computing $\langle \varLambda^i,\,$ $ \{ A^i_j \} _{j=1}^{l}, \, \{ B^i_j \} _{j=1}^{m_i} \rangle$  for  $1\leq i \leq h$ has  $\mathcal{O}(h\tilde{l}^4+ h\tilde{l}^3s + h\tilde{l}^2s^2 )$ complexity. The complexity of all other operations is a lower-order term, and hence the overall complexity is  $\mathcal{O}(h\tilde{l}^4+ h\tilde{l}^3s + h\tilde{l}^2s^2 )$.
 	\end{Pro}

\subsection{Set Validations} \label{section: Set Validations}

In this section, we  introduce  the necessary set validations  in SE, RA, and FD for spectrahedral shadows. \vspace{1pt}
   
  \subsubsection{Emptiness Check}
  For RA and FD, it is common to test whether two sets  intersect (i.e., the intersection of the sets is empty), such as collision check \cite{kousik2023} and    set separation \cite{xu2022b}.   Proposition \ref{emptyness check propos} shows how to check the emptiness for spectrahedral shadows.
   \begin{Propos}[Emptiness Check]   \label{emptyness check propos}
  	For arbitrary $ \mathcal{S} =  \langle \varLambda,$  $\, \{ A_i \}_{i=1}^{n},\, \{ B_i \} _{i=1}^{m} \rangle $ with size $s$, $ \mathcal{S} = \emptyset$ iff $\epsilon_e^* < 0$, and $\epsilon^*$ is obtained by solving
  	   \begin{equation} \label{emptyness check problem}
  		\begin{aligned}
  		\epsilon_e^* =  \max_{\epsilon,\,x,\,y} \; \epsilon  \quad s.t., \varLambda  + \sum_{i=1}^{n} x_i A_i +\sum_{i=1}^m{y_iB_i}\succeq  \epsilon I,
  		\end{aligned}
  	\end{equation}
  	where $x\in \mathbb{R}^n$ and $y\in \mathbb{R}^m$. The complexity to test whether $ \mathcal{S} $ is an empty set is $\mathcal{O}(\sqrt{s}((n+m)s^2+(n+m)^\omega+s^\omega))$. \vspace{3pt}
  \end{Propos}
  
  \begin{Pro}
    Let $\varLambda(x,\,y) =  \varLambda  + \sum_{i=1}^{n} x_i A_i +\sum_{i=1}^m{y_iB_i}$. According to Definition \ref{def of spectrahedral shadows},  $ \mathcal{S} = \emptyset  $ iff $ \nexists \, x \in \mathbb{R}^n, \,  y \in \mathbb{R}^m, \,  \varLambda(x,\,y) \succeq 0$, which is further equivalent to $\max_{x,\,y}\; \lambda_{\min}(\varLambda(x,\,y) )  < 0$. Then $\max_{x,\,y}\, \lambda_{\min}(\varLambda(x,\,y))$ is obtained by the well-known   method to minimize  the spectral norm, i.e., solving  the SDP problem \eqref{emptyness check problem}.
      
     \textit{Complexity:} Solving the SDP problem \eqref{emptyness check problem} runs in  $\mathcal{O}(\sqrt{s}((n+m+1)s^2+(n+m+1)^\omega+s^\omega))$ time. The complexity of comparing $\epsilon^*$ with $0$ is $\mathcal{O}(1)$. By omitting the lower-order term,  the overall complexity is $\mathcal{O}(\sqrt{s}((n+m)s^2+(n+m)^\omega+s^\omega))$.
     \end{Pro}

 \subsubsection{Point Containment}
 Point containment is a basic task in FD, since it  needs to check whether the real system output lies in the output set of a healthy system \cite{xu2019b}. We thus introduce Corollary \ref{Point Containment propos}  to  validate this point  for spectrahedral shadows.
 
  \begin{Coro}[Point Containment]   \label{Point Containment propos}
 	For arbitrary  $v \in \mathbb{R}^n$ and $ \mathcal{S} =  \langle \varLambda,$  $\, \{ A_i \}_{i=1}^{n},\, \{ B_i \} _{i=1}^{m} \rangle $ with size $s$, $ v \in \mathcal{S}$ iff $\epsilon_p^* \geq 0$, and $\epsilon^*$ is obtained by solving 
 	\begin{equation} \label{Point Containment problem}
 		\begin{aligned}
 			\epsilon_p^* =  \max_{\epsilon,\,y} \; \epsilon  \quad s.t., \varLambda^v   +\sum_{i=1}^m{y_iB_i}\succeq  \epsilon I,
 		\end{aligned}
 	\end{equation}
 	where $\varLambda^v   = \varLambda  +\sum_{i=1}^{n} v_i A_i$ and $y\in \mathbb{R}^m$. The complexity to test whether $x_t$ lies in $ \mathcal{S} $  is $\mathcal{O}(ns^2+\sqrt{s}(ms^2+m^\omega+s^\omega))$. \vspace{3pt}
 \end{Coro}
 
   \begin{Pro}
    Corollary \ref{Point Containment propos}  follows from Proposition  \ref{emptyness check propos} by noticing that $ v \in \mathcal{S} \Leftrightarrow  \langle \varLambda^v, \{ B_i \} _{i=1}^{m} \rangle \neq \emptyset $ and the complexity of computing $\varLambda^v $ is $\mathcal{O}(2ns^2)$.
   \end{Pro}

  \subsubsection{Boundedness Check}

   In practice, not all sets are bounded, such as  consistent state
   sets \cite{alamo2005} and  unguaranteed detectable faults sets \cite{wang2024}. Thus,    a numerical method is proposed to check the boundedness of spectrahedral shadows.
   
    \begin{Propos}[Boundedness Check]  \label{Boundedness check propos}
  	For any nonempty  $ \mathcal{S} =  \langle \varLambda,$  $\, \{ A_i \}_{i=1}^{n},\, \{ B_i \} _{i=1}^{m} \rangle $ with size $s$, $\mathcal{S}$ is bounded iff  \vspace{1.5pt}  $\mathrm{rank}(P) = n$, $\mathrm{rank}(P) + \mathrm{rank}(Q) =  \mathrm{rank}([P,\, Q])$ and either of the following conditions holds
  	\begin{enumerate} 
  		\item    $\mathrm{tr}(A_i) = 0, \, \forall \; 1\leq i \leq n$ and  $\mathrm{tr}(B_j) = 0, \, \forall \; 1\leq j \leq m$; 
  		\item   $\epsilon_b^* < 0$,
  	\end{enumerate}
  	  where $ P =  [\, \mathrm{tvec}(A_1),\,$ $ \mathrm{tvec}(A_2), \,  ...,  \mathrm{tvec}(A_n) ]\in \mathbb{R}^{{s(s+1)/2}\times n}$, $ Q =  [\, \mathrm{tvec}(B_1),\,$ $ \mathrm{tvec}(B_2), \,  ...,  \mathrm{tvec}(B_n) ] \in \mathbb{R}^{{s(s+1)/2}\times m} $,
  	and  $\epsilon_b^* $ is obtained by solving 
  	\begin{subequations} \label{Boundedness check problem}
  		\begin{align}
  			&\epsilon_b^* =  \max_{\epsilon,\,x,\, y} \; \epsilon  \quad \\ 
  			& s.t.,  \sum_{i=1}^{n} x_i A_i +\sum_{i=1}^{m} y_i B_i  \succeq  \epsilon I, \\
  			& \quad \;\; \sum_{i=1}^{n} \mathrm{tr}(A_i) x_i +\sum_{i=1}^{m} \mathrm{tr}(B_i) y_i = 1.
  		\end{align}
  	\end{subequations}
  	The complexity to test whether  $\mathcal{S}$ is bounded is $\mathcal{O}((m+n)s^4+\sqrt{s}(m+n)^\omega)$. \vspace{3pt}
  \end{Propos}
  
  \begin{Pro}
   We first prove that  $\mathcal{S}$ is unbounded iff  $\exists\, u \in \mathbb{R}^{n} \backslash \{\textbf{0}\} , \, v \in \mathbb{R}^{m}, \; \sum_{i=1}^{n} u_i A_i + \sum_{i=1}^{m} v_i B_i  \succeq  0$. The sufficiency is clear by  verifying $ \lim_{\alpha \rightarrow \infty} x+ \alpha u \in \mathcal{S}$ for $ \forall \, x \in  \mathcal{S}$, where $ \alpha \in \mathbb{R}$. For  the necessity,  consider  arbitrary  $x  \in \mathcal{S}$.  Since $\mathcal{S}$ is unbounded, there exists   $d\in \mathbb{R}^{n}\backslash \{\textbf{0}\}$ such that $x^\alpha = x + \alpha d \in  \mathcal{S} $ holds for all $\alpha \geq 0$. Let  $u^\alpha =  \frac{x^\alpha}{\|x^\alpha\| } $ and $u^{\infty} = \lim_{\alpha \rightarrow +\infty} u^\alpha$. We have 
   	   \begin{equation} 
   	\begin{aligned}
  \exists\, y\in \mathbb{R} ^m,\,	\frac{1}{\|x^\alpha \| } \varLambda  + \sum_{i=1}^{n}  u^\alpha_i A_i +\sum_{i=1}^m{  \frac{y_i}{\|x^\alpha\| }  B_i}\succeq  0,
   	\end{aligned}
   \end{equation}
   which yields $\exists\, v =  \frac{y}{\|x^\alpha\| }, \, \sum_{i=1}^{n} u^{\infty}_i A_i + \sum_{i=1}^{m} v_i B_i  \succeq  0$. Since $u^\alpha \in \mathscr{S}=\{x\in \mathbb{R} ^n\! : \|x\|= 1 \} $  and  $ \mathscr{S}$ is a closed set, we have $u^{\infty} \in  \mathscr{S} \subset \mathbb{R}^{n} \backslash \{\textbf{0}\} $, thereby leading to the necessity.
   
   Then consider Proposition \ref{Boundedness check propos} in three  cases. For the case of  $\mathrm{rank}(P) < n$, we know  $\exists \, u^* \in \mathbb{R}^{n} \backslash \{\textbf{0}\}, \,  P u^* = \textbf{0} $, which is equivalent to $\sum_{i=1}^{n} u^*_i A_i = \textbf{0} $. Since $\exists\, v =  \textbf{0}, \, \sum_{i=1}^{n} u^{*}_i A_i + \sum_{i=1}^{m} v_i B_i  \succeq  0$, $\mathcal{S}$ is unbounded.   
   
   For the second case, i.e., $\mathrm{rank}(P) = n$ and  $\mathrm{rank}(P) + \mathrm{rank}(Q) <  \mathrm{rank}([P,\, Q])$, let  $\mathcal{R}_P = \{\sum_{i=1}^{n} x_i P_{[i,*]} \! : x\in \mathbb{R}^n \}$ and $\mathcal{R}_Q = \{\sum_{i=1}^{m} x_i Q_{[i,*]} \! : x\in \mathbb{R}^m \}$  be the \textit{range} of $P$ and $Q$, respectively.  Then  $\mathrm{rank}(P) + \mathrm{rank}(Q) <  \mathrm{rank}([P,\, Q])$ is equivalent to  $\mathrm{dim}(\mathcal{R}_P) + \mathrm{dim}(\mathcal{R}_Q) <  \mathrm{dim}(\mathcal{R}_P \oplus \mathcal{R}_Q)$. By subspace intersection lemma \cite[(0.1.7.1)]{horn1985}, we have $ \mathrm{dim}(\mathcal{R}_P \cap \mathcal{R}_Q) = \mathrm{dim}(\mathcal{R}_P \oplus \mathcal{R}_Q) - (\mathrm{dim}(\mathcal{R}_P) + \mathrm{dim}(\mathcal{R}_Q)) > 0$, i.e.,  $\mathcal{R}_P \cap \mathcal{R}_Q  \varsupsetneqq \{ \textbf{0} \}$. Since $P$ is full column rank, $\mathcal{R}_P \cap \mathcal{R}_Q \varsupsetneqq \{ \textbf{0} \} $ iff $  \exists\, x \in \mathbb{R}^{n} \backslash \{\textbf{0}\},\, y\in  \mathbb{R}^{n},\, \sum_{i=1}^{n} x_i P_{[i,*]} = \sum_{i=1}^{n} y_i Q_{[i,*]}$. Let $u = x$ and $v = -y$. We have $\sum_{i=1}^{n} u_i A_i + \sum_{i=1}^{m} v_i B_i = \textbf{0} \succeq 0$, and hence $\mathcal{S}$ is unbounded.

   For the third case, i.e.,  $\mathrm{rank}(Q) = n$ and $\mathrm{rank}(P) + \mathrm{rank}(Q) =  \mathrm{rank}([P,\, Q])$, it can be concluded that $\nexists \, u \in \mathbb{R}^{n} \backslash \{\textbf{0}\} , \, v \in \mathbb{R}^{m}, \; \sum_{i=1}^{n} u_i A_i + \sum_{i=1}^{m} v_i B_i = \textbf{0}$ by analogy  with the second case. 
    Consider    the subspace $\mathcal{R}_{AB} = \{\sum_{i=1}^{n} x_i A_i + \sum_{i=1}^{m} y_i B_i\!: x\in \mathbb{R}^n, \, y\in \mathbb{R}^m \}$ and denote $\mathbb{S}^{n}_{+} = \{X\in \mathbb{S}^{n} \!: X\succeq 0 \}$. In this case,
   $\mathcal{S}$ is unbounded (i.e., $\exists\, u \in \mathbb{R}^{n} \backslash \{\textbf{0}\} , \, v \in \mathbb{R}^{m}, \; \sum_{i=1}^{n} u_i A_i + \sum_{i=1}^{m} v_i B_i  \succeq  0$) iff  $\mathcal{R}_{AB} \cap  \mathbb{S}^{n}_{+}  \varsupsetneqq \{ \textbf{0} \} $. Notice that both $\mathcal{R}_{AB}$ and  $\mathbb{S}^{n}_{+}$ are cones in $\mathbb{S}^n$,  $\mathcal{R}_{AB}  \cap  \mathbb{S}^{n}_{+}$ is  a cone  in $\mathbb{S}^{n}_{+}$,  and hence
   \begin{equation} \label{boundness tmp}
   	\begin{aligned}
   	\mathcal{R}_{AB}\cap  \mathbb{S}^{n}_{+}  \varsupsetneqq \{ \textbf{0} \} &\Leftrightarrow  	(	\mathcal{R}_{AB}\cap  \mathbb{S}^{n}_{+} )\cap  \mathrm{spx}(n)  \neq  \emptyset \\& \Leftrightarrow \mathcal{R}_{AB}\cap  \mathrm{spx}(n)  \neq  \emptyset,
   	\end{aligned}
   \end{equation}
   where $\mathrm{spx}(n) = \{X\in \mathbb{S}^{n}_{+}\!: \mathrm{tr}(X) = 1 \}$ is the \textit{spectraplex} (i.e., the generalization of simplex in $\mathbb{S}^{n}$).  Then $\mathcal{R}_{AB}\cap  \mathrm{spx}(n)  \neq  \emptyset$ can be validated by  analogy  with Proposition \ref{emptyness check propos}, thereby leading to  Proposition \ref{Boundedness check propos} (Note that the problem \eqref{Boundedness check problem} is infeasible when $\mathrm{tr}(A^i) = 0, \, \forall \; 1\leq i \leq n$ and  $\mathrm{tr}(B_j) = 0, \, \forall \; 1\leq j \leq m$,   and hence $\mathcal{R}_{AB}\cap  \mathrm{spx}(n)  =  \emptyset$).
   
  \textit{Complexity:}  When $s(s+1)/2 < n $,  $\mathcal{S}$  is clearly unbounded  due to $\mathrm{rank}(P)  < n$. When $s(s+1)/2 \geq n $, the complexity to compute $\mathrm{rank}(P)$, $\mathrm{rank}(Q)$ and $\mathrm{rank}([P, \, Q])$ using SVD is $\mathcal{O}(2(m+n)s^4)$, and the complexity to solve the SDP problem \eqref{Boundedness check problem} is $\mathcal{O}(\sqrt{s}((m+n)s^2+(m+n)^\omega+s^\omega))$. Since the complexity of all other operations is a lower-order term, the overall complexity is reduced as $\mathcal{O}((m+n)s^4+\sqrt{s}(m+n)^\omega)$ by recalling $\omega \leq 2.372$.
  \end{Pro}

\subsection{Conversion from Existing Set Representations} \label{section: relation with existing set representation}
In this section, we  discuss  some existing convex set representations that can be formulated as spectrahedral shadows.\vspace{3pt}

\subsubsection{H-polyhedron}  We first show how to convert  H-polyhedrons into spectrahedrons. \vspace{3pt} 

\begin{Lem}[Conversion from H-polyhedrons]   \label{propos of H-polyhedron}
  Given an H-polyhedron	$\mathcal{P}(A, b)=\{x \in \mathbb{R}^{n}\! : Ax\leq b\}$ with $A \in \mathbb{R}^{m\times n}$ and $b \in \mathbb{R}^{m}$, $\mathcal{P}$ can be represented by a spectrahedron  $ \langle \varLambda , \, \{ A_i \} _{i=1}^{n} \rangle  $ with size $m$, where $\varLambda = \mathrm{diag}(b_1,b_2,...,b_m) $ and $A_i = \mathrm{diag}(-A_{[1,\, i]},\, -A_{[2,\, i]},...,\,-A_{[m,\, i]}), \, \forall \, 1\leq i \leq n $.\vspace{3pt} 
\end{Lem}

\begin{Pro}
 	It is clear to formulate 	$\mathcal{P}(A, b)$ as $ \langle \varLambda , \, \{ A_i \} _{i=1}^{n} \rangle  $ according to Definition \ref{def of spectrahedral shadows}.
 \end{Pro}

\subsubsection{Ellipsoid} Ellipsoids can also be readily represented as spectrahedral shadows.\vspace{3pt}

\begin{Lem}[Conversion from Ellipsoids]   \label{propos of Ellipsoid}
	Given an ellipsoid $\mathcal{E}(c,\, Q) = \{x \in \mathbb{R}^{n}: (x-c)^TQ^{-1}(x-c) \leq 1,\, Q\succ 0 \}$ with $Q\in \mathbb{S}^{ n}$ and $c \in \mathbb{R}^{n}$, $\mathcal{E}(c,\, Q)$ can be represented by a spectrahedron  $ \langle \varLambda , \, \{ A_i \} _{i=1}^{n} \rangle  $ with size $n+1$, where
	\begin{equation} \notag
			\resizebox{1.0\hsize}{!}{
			$		\varLambda \hspace{-0.5pt} = \hspace{-0.5pt} \left[ \begin{matrix}
			\,Q  \hspace{-6pt}   &	-c  \,  \\
			\,-c ^T \hspace{-6pt}  &	1  \
		\end{matrix} \right]\! ,\;  A_{\left[ j,k \right]}^{i}=  \hspace{-0.5pt} \begin{cases}
		1, \hspace{6pt}   j=n+i, \,k=1\\
		1,  \hspace{6pt}  j=1,\, k=n+i\\
		0, \hspace{6pt}  \text{otherwise}\\
		\end{cases} \hspace{-12pt}, \, \forall \, 1 \leq i \leq n.
		$}
	\end{equation}\vspace{1pt} 
\end{Lem}

\begin{Pro}
By Schur complement lemma, $(x-c)^TQ^{-1}(x-c) \leq 1 $ and $Q\succ 0$ iff  $\left[ \begin{matrix}
 	\,Q  \hspace{-6pt}   &	x-c  \,  \\
 	\,(x-c) ^T \hspace{-6pt}  &	1  \
 \end{matrix} \right] \succeq 0$, which yields Proposition \ref{propos of Ellipsoid}.
\end{Pro}

\subsubsection{Zonotope}  Then we show how to convert zonotopes into  spectrahedral shadows based on the developed set operations.\vspace{3pt}

\begin{Propos}[Conversion from Zonotopes]   \label{propos of zonotope}
	Given a zonotope $\mathcal{Z}(c,\, G) = \{ x \in \mathbb{R}^{n}: x=c+G\xi,\, \left\| \xi \right\|_\infty \leq 1 \}$ with  $c \in \mathbb{R}^{n}$   and $G\in \mathbb{R}^{ n \times n_g}$,  $\mathcal{Z}(c,\, G)$ can be represented by a spectrahedral shadow  $ \langle \varLambda , \, \{ A_i \} _{i=1}^{n} , \, \{ B_i \} _{i=1}^{n_g-r_g} \rangle  $ with size $2(n+n_g-r_g)$, where  $r_g = \mathrm{rank}(G)$, $G=UDV^T$ is the  SVD of $G$, $G^{\dagger}$ is the  Moore-Penrose inverse  of $G$, and
	\begin{equation} \notag
	\begin{aligned}
	 &\varLambda = \mathrm{diag}(I_{2n_g}, \textbf{0}) - \sum_{i=1}^{n}c_i A_i, \\
	 &A_i = \mathrm{diag}(G^{\dagger}_{[1, \, i]},\, G^{\dagger}_{[2, \, i]} ,..., G^{\dagger}_{[n_g, \, i]},\,  -G^{\dagger}_{[1, \, i]},\, -G^{\dagger}_{[2, \, i]} ,...,\\
	 & \qquad \quad \; \, -G^{\dagger}_{[n_g, \, i]}, \, U_{[r_g+1, \, i]} , \, U_{[r_g+2, \, i]}, ..., U_{[n, \, i]}, \\ & \qquad \quad \; \,- U_{[r_g+1, \, i]} , \,- U_{[r_g+2, \, i]}, ..., -U_{[n, \, i]}   ),\, \forall \, 1\leq i \leq n, \\
	 & B_i  =  \mathrm{diag}(V_{[1, \, r_g+i]},\, V_{[2, \, r_g+i]} ,..., V_{[n_g, \, r_g+i]}, \, -  V_{[1, \, r_g+i]}, \\
	 & \qquad \quad \; \, -   V_{[2, \, r_g+i]}..., -V_{[n_g, \, r_g+i]}) ,\, \forall \, 1\leq i \leq n_g-r_g.\\ \noalign{\vskip 10pt}
	\end{aligned}
	\end{equation}
\end{Propos}

\begin{Pro}
	$\mathcal{Z}(c,\, G)$ can be rewritten as  $c + G \circ \mathcal{B}_{\infty}^{\hspace{0.5pt} n_g}$, where the unit $\infty$-norm ball $\mathcal{B}_{\infty}^{\hspace{0.5pt}n_g}$ is represented by the    H-polyhedron  $\{x \in \mathbb{R}^{n_g}\! : [I_{n_g},\, -I_{n_g}]^T x\leq \textbf{1}\}$. Using Lemma \ref{propos of H-polyhedron},  $\mathcal{B}_{\infty}^{\hspace{0.5pt}n_g}$ can be further formulated as a spectrahedral shadow. Due to 	$\mathcal{Z}(c,\, G)= c + G \circ \mathcal{B}_{\infty}^{\hspace{0.5pt}n_g}$, $\mathcal{Z}(c,\, G)$ is formulated as a spectrahedral shadow   by Propositions \ref{propos of linear  map} and  \ref{propos of translation}. 
\end{Pro}

\subsubsection{Ellipsotope  \& Constrained Zonotope} For more complex set representations like ellipsotopes, we  provide an algorithm to convert  them into  spectrahedral shadows.  This algorithm is equally applicable for constrained zonotopes, as they are readily written as ellipsotopes \cite[Section \uppercase\expandafter{\romannumeral 4}-E]{kousik2023}  

The basic idea of the algorithm is   regarding ellipsotopes as a combination of  some spectrahedral shadows under the translation, linear map and intersections. Specifically, for arbitrary $\mathcal{E}_p(c,\, G, \, A, \, b, \, \mathscr{J}) = \{ x \in \mathbb{R}^{n}\!:  x=c+G\xi,\,  \left\| \xi\langle J \rangle \right\|_p \leq 1, \, \forall \, J \in \mathscr{J}, \, Ax =b \}$ with $\mathscr{J}= \{J_1, J_2, ..., J_{|\mathscr{J}|}  \} \subset \mathbb{P}(\mathbb{N}) $, we have
   \begin{equation} \notag
	\begin{aligned}
	\mathcal{E}_p(c,\, G, \, A, \, b, \, \mathscr{J}) = c + G \circ ( \mathcal{P}(A, b) \cap   \mathcal{P}(A, -b)\cap  \mathcal{B}^{\times}_p(\mathscr{J})),
	\end{aligned}
\end{equation}
where $\mathcal{B}^{\times}_p(\mathscr{J}) = \{ x \in \mathbb{R}^{n}\!:  \left\| \xi\langle J \rangle \right\|_p \leq 1, \, \forall \, J \in \mathscr{J} \}$. In \cite{kousik2023}, $\mathcal{B}^{\times}_p(\mathscr{J})$ is referred to  as  a \textit{ball product}, which is equivalently written as  
   \begin{equation} \label{ball product}
	\begin{aligned}
	\mathcal{B}^{\times}_p(\mathscr{J}) =  \mathcal{B}_{p}^{|J_1|}\times  \mathcal{B}_{p}^{|J_2|} \times \cdots \times \mathcal{B}_{p}^{|J_{|\mathscr{J}|}|} .
	\end{aligned}
\end{equation}
It has been shown in \cite[(3.3.37)]{ben2001} that any $\mathcal{B}_{p}^n$ with rational $p$ can be formulated as  conic quadratic-representable sets,  and such sets can be further formulated as spectrahedral shadows \cite[(4.2.1)]{ben2001}. Since all required set operations have been given in Section \ref{section: set operations}, it yields Algorithm \ref{Conversion from ellipsotope}.

\begin{algorithm}[h] 
	\caption{\hspace{-3pt}\textbf{:}$\;$Conversion from Ellipsotopes}
	\label{Conversion from ellipsotope}
	\begin{algorithmic}[1]
		\Require  $\mathcal{E}_p(c,\, G, \, A, \, b, \, \mathscr{J}) $ with   $\mathscr{J}= \{J_1, J_2, ..., J_{|\mathscr{J}|}  \}$;
		\Ensure  the spectrahedral shadow $ \langle \varLambda , \, \{ A_i \} _{i=1}^{n},\, \{ B_i \} _{i=1}^{m} \rangle  $;
		\State  Formulate $\mathcal{B}_{p}^{|J_1|}$,  $\mathcal{B}_{p}^{|J_2|} $,.., \,$ \mathcal{B}_{p}^{|J_{|\mathscr{J}|}|}$ as spectrahedral shadows using (3.3.37) and (4.2.1) in \cite{ben2001} \vspace{0.8pt};
		\State Compute $	\mathcal{B}^{\times}_p(\mathscr{J})$ according to \eqref{ball product} and Proposition \ref{propos of Cartesian product};
		\State Formulate $\mathcal{P}(A, b)$ and $\mathcal{P}(A, -b)$ as spectrahedral shadows using Lemma \ref{propos of H-polyhedron};
		\State Compute $\mathcal{S}_e=\mathcal{P}(A, b) \cap   \mathcal{P}(A, -b)\cap  \mathcal{B}^{\times}_p(\mathscr{J})$ using Lemma \ref{lemma of intersection};
		\State Compute $  \langle \varLambda , \, \{ A_i \} _{i=1}^{n},\, \{ B_i \} _{i=1}^{m} \rangle = c + G \circ  \mathcal{S}_e$ using Propositions \ref{propos of linear  map} and \ref{propos of translation};
	\end{algorithmic}
\end{algorithm}

\section{Complexity Reduction Strategy} \label{section: Complexity Reduction Strategy}

For spectrahedral shadows, the  growth of size and lifted dimension is a realistic issue  when  iteratively implementing set operations like the Minkowski sum. This issue  commonly exists in  SE, RA and FD \cite{raghuraman2022, rego2025}, resulting in the increase of time and space overhead for subsequent set operations. Thus, 
reducing the complexity of spectrahedral shadows is necessary for many set-based applications.

In this section,  we first propose a method called  \textit{local low-rank approximation} to compute a lower size spectrahedron as the outer approximation of a given spectrahedron.  Then we discuss how to reduce  the size and eliminate the lifted dimensions for general spectrahedral shadows. Finally, we provide strategies to   accelerate the above complexity-reducing process in practice by utilizing the sparsity of    spectrahedral shadows. 

\subsection{Order Reduction of Spectrahedrons}  \label{section: order reduction for spectrahedrons}

\begin{figure}[!t]
	\centering
	\subfigure{
		\hspace{-7pt} \includegraphics[scale=0.21]{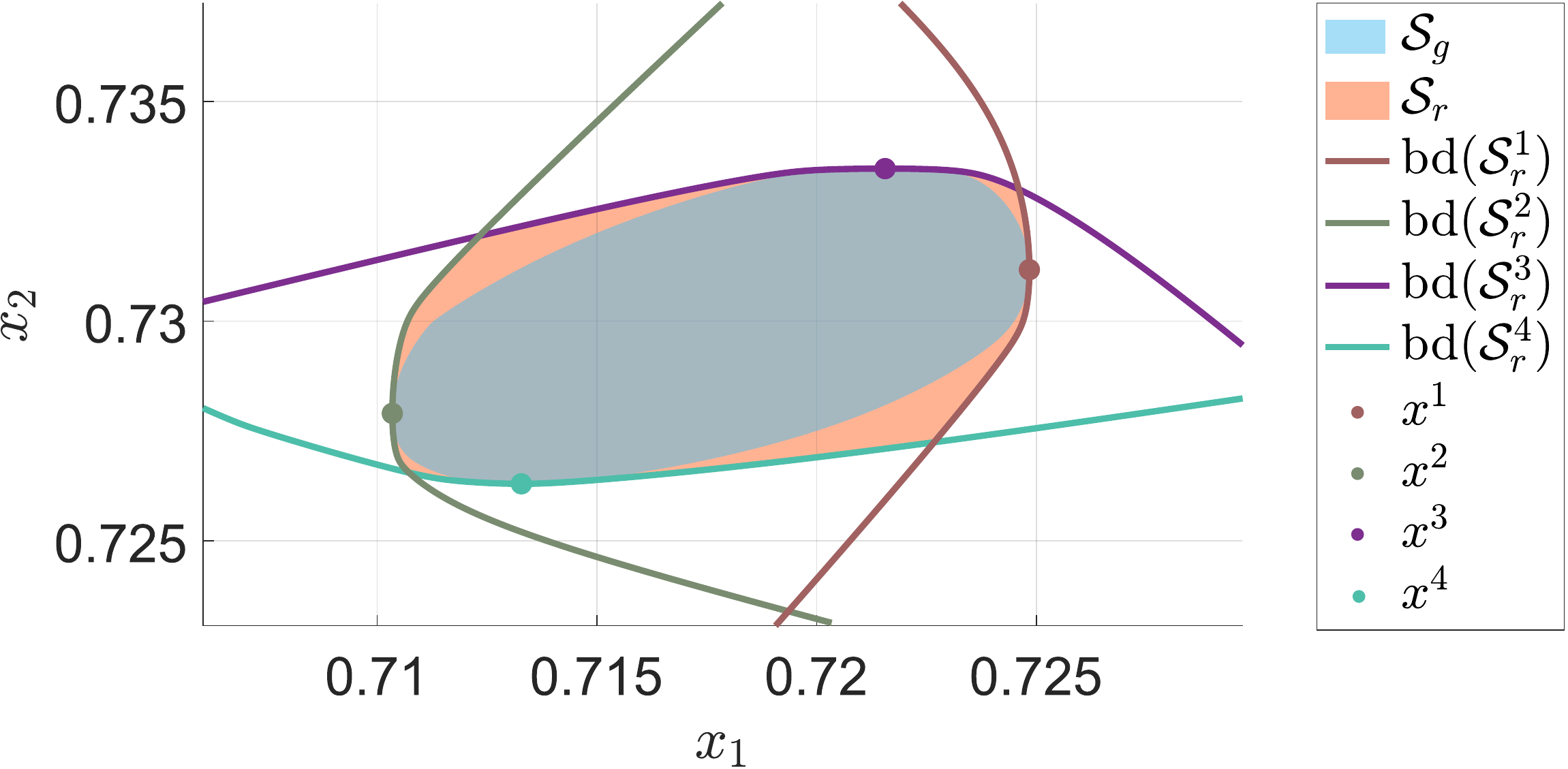}
	}
	\vspace{-4pt}
	\caption{Example of local low rank approximation. $\mathcal{S}_g$ is a spectrahedron to be reduced size ($s_g = 50$). $\mathcal{S}_r = \mathcal{S}_r^1 \cap \mathcal{S}_r^{2} \cap \mathcal{S}_r^{3} \cap\mathcal{S}_r^{4}$ is the reduced-size spectrahedron  ($s_r = 40$). $\text{bd}{(\mathcal{S}_r^i)}$ ($1\leq i \leq 4 $) is  the part of the boundary of $\mathcal{S}_r^i$ that locally approximates $\text{bd}{(\mathcal{S}_g)}$  at $x^i$.} \label{fig6}
\end{figure}

 For brevity, we use $\varLambda(x) =  \varLambda +\sum_{i=1}^n{x_iA_i}$ and $\varLambda(x,y) = \varLambda +\sum_{i=1}^n{x_iA_i}+\sum_{i=1}^m{y_iB_i}$ to denote the linear matrix pencils in this section. Assume that $\mathcal{S}_g = \{x \in \mathbb{R}^{n} \! : \varLambda(x) \succeq 0 \} $ is a full-dimensional bounded spectrahedron to be reduced size with  size $s_g$ and   there are $N$ different points $x^1, x^2,...,x^{N}$  on the boundary of $\mathcal{S}_g $ (i.e., $x_i \in \mathrm{bd}(\mathcal{S}_g)$). The  idea of the local low-rank approximation is to produce a reduced-size spectrahedron $\mathcal{S}_r = \mathcal{S}_r^1 \cap \mathcal{S}_r^{2} \cap.... \cap\mathcal{S}_r^{N}$ with the size $s_r$ such that:
 \begin{enumerate}
 	\item   $\mathcal{S}_g \subseteq  \mathcal{S}_r^i, \, \forall\, 1\leq i \leq N$;
 	\item   Each $\mathcal{S}_r^i$  locally approximates $\mathcal{S}_g$ at the point $x^i$. Namely, $\mathcal{S}_g$ is internally tangent to $\mathcal{S}_r^i$  at  $x^i$. \;\;
 \end{enumerate}
 Fig. \ref{fig6} gives an example to illustrate the above process.

Next, we  elaborate  on the proposed method. Consider that there is a matrix $P^i \in \mathbb{R}^{s_g \times s_r^i}$.  Each $\mathcal{S}_r^i$ is computed based on the following implication:
\begin{equation} \label{implication relation}
	\begin{aligned} 
	\varLambda(x) \succeq 0  \Rightarrow   (P^i)^T	\varLambda(x) P^i \succeq 0 .
	\end{aligned}
\end{equation}
Let $\mathcal{S}_r^i =  \{x \in \mathbb{R}^{n} \! :    (P^i)^T	\varLambda(x) P^i \succeq 0 \}$. Then we have $\mathcal{S}_g \subseteq  \mathcal{S}_r^i$ for arbitrary  $P^i$. \vspace{1pt}

 Then we  show how to choose the proper $  P^i$ such that   $\mathcal{S}_r^i$  locally approximates $\mathcal{S}_g$ at $x^i$. Specifically, assume that the SVD  of $	\varLambda(x^i)$  is   $	\varLambda(x^i) = U^i D^i (U^i)^T$, where $D^i$ is a diagonal matrix such that $D^i_{[j,j]}$ ($1\leq j \leq s_g$) is the $j$-th largest singular value of  $	\varLambda(x^i)$. Then $  P^i$ is given by 
   \begin{equation} 
 	\begin{aligned}
 	P^i = [\,U^i_{[*,\, s_g-s_r+1]}, \, U^i_{[*, \, s_g-s_r+2]}, ..., U^i_{[*, \,s_g]}\, ].
 	\end{aligned}
 \end{equation}
Since the necessary and sufficient condition  for $x \in \mathrm{bd}(\mathcal{S}_g)$  is  $\det(\varLambda(x)) = 0$ \cite[Theorem 3.2.3]{sletsjoe2023}, we can prove that $x^i \in \mathrm{bd}(\mathcal{S}_g) $ implies $ x^i  \in \mathrm{bd}(\mathcal{S}_r^i) $   with such a choice of $  P^i$. Due to  $\mathcal{S}_g \subseteq  \mathcal{S}_r^i$ and  $ x^i  \in \mathrm{bd}(\mathcal{S}_r^i) \cap   \mathrm{bd}(\mathcal{S}_g)$,  $\mathcal{S}_g$ is internally tangent to $\mathcal{S}_r^i$  at  $x^i$. Moreover,
  it is clear that:
 \begin{enumerate}
 	\item   When $\mathrm{rank}(P^i ) \geq s_r $,   $\mathcal{S}_r^i$ is exactly  $\mathcal{S}_g$;
 	\item    When $\mathrm{rank}(P^i ) = 1 $, $\mathcal{S}_r^i$ is a half-space and $\text{bd}{(\mathcal{S}_r^i)}$ is the  hyperplane tangent to $\mathcal{S}_g$  at  $x^i$. \;\;
 \end{enumerate}
 For $1 < \mathrm{rank}(P^i ) < s_r$, our numerical experiments show that with the increase of $\mathrm{rank}(P^i )$, the approximation of  $\mathcal{S}_r^i$  to $\mathcal{S}_g$ becomes more accurate.

 The remaining issue is the choice of boundary points  $x^1, x^2,...,x^{N}$. After experimenting with various strategies, an efficient heuristic is found to be choosing $N = 2n$ and
    \begin{equation} \label{boundary point sdp}
 	\begin{aligned}
 		x^i =  \begin{cases}
 		\, \arg \max_{x\in \mathcal{S}_g }\, x_i, \;\,  1\leq i \leq n	\\
 		\, \arg \min_{x\in \mathcal{S}_g }\, x_{i-n}, \;\,  n+1\leq i \leq 2n	\\
 		\end{cases}
 	\end{aligned}
 \end{equation}
  if the shape of  $\mathcal{S}_g $ is approximately  isotropic  (e.g., approximates a unit ball or a hypercube). In practice, since  $\mathcal{S}_g $ does not always approximate to an isotropic set (e.g.,  the shape of $\mathcal{S}_g $ may be long and narrow), we can find a nonsingular transformation matrix $T^* \in  \mathbb{R}^{n \times n} $ such that $\mathcal{S}_g^{'} = T^*  \circ \mathcal{S}_g $  is an isotropic set (see Appendix \ref{The computation of T} for the details). By applying the above process,   a reduced-size spectrahedron $\mathcal{S}_r^{'}$  is computed for $\mathcal{S}_g^{'}$. Thus,  $\mathcal{S}_r =   \mathcal{S}_r^{'} \circ T^* $ is the reduced-size spectrahedron for $\mathcal{S}_g$.
  
 By properly choosing $s_r^i$ such that $s_r =\sum_{i=1}^{N} s_r^i\leq s_g$, the above process leads to a reduced-size spectrahedron $\mathcal{S}_r$. 
 
\begin{figure}[!t]
	\centering
	\subfigure{
		\hspace{-7pt} \includegraphics[scale=0.29]{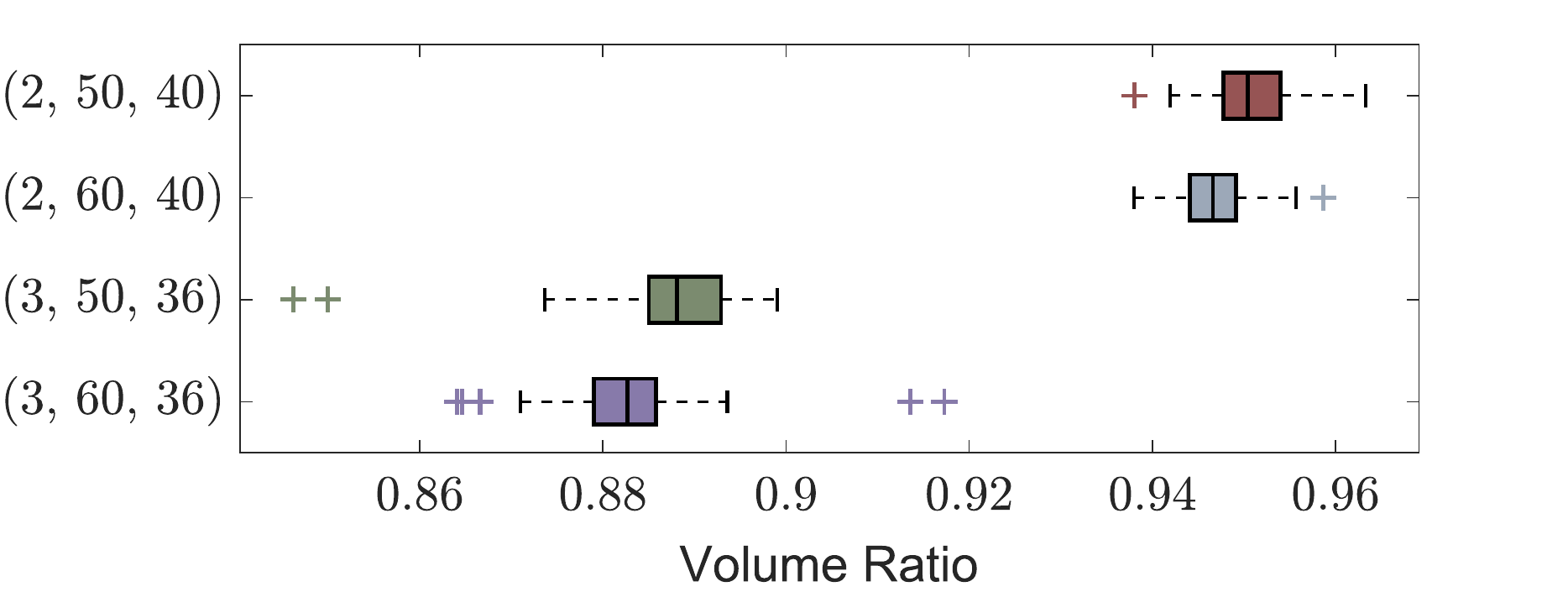}
	}
	\vspace{-4pt}
	\caption{The box plot for the volume ratio $\Delta V$ of   $\mathcal{S}_g$ and  $\mathcal{S}_r$. Each row  label $(n\,, s_g,\, s_r)$ corresponds to  a group of results for $100$ random simulations, where the  spectrahedron $\mathcal{S}_g \subset \mathbb{R}^n$ with the size $s_g$ is outer approximated by the  spectrahedron  $\mathcal{S}_r  \subset \mathbb{R}^n$  with the size $s_r$.} \label{fig7}
\end{figure}

 \begin{Exam}
 	To assess the effectiveness, Fig. \ref{fig7} shows the box plot for the volume ratio\footnote{In  this paper, the volume computation for spectrahedral shadows  are computed by high precision boundary sampling, as there is currently no analytical expression for computing the volume of   spectrahedral shadows.} of  $\mathcal{S}_g$ and  $\mathcal{S}_r$ over 4 groups of  examples. In each group, we randomly generate  100 $n$-dimensional $\mathcal{S}_g$, and compute $\mathcal{S}_r $ with the size $s_r$ by the proposed method, where $s_r^i$ ($1 \leq i \leq 2n$) is chosen as $s_r^i = \frac{s_r}{2n} $. In Fig. \ref{fig7}, it is observed  that the  reduced-size  spectrahedron $\mathcal{S}_r$ efficiently approximates  $\mathcal{S}_g$, and the approximation becomes more accurate with the decrease  in the size to be reduced.
 \end{Exam}

\subsection{Polyhedral Approximation of Spectrahedral Shadows}  \label{section: Polyhedral Approximation}
 It is tricky to find a general spectrahedral shadow with smaller size and lifted dimensions to outer approximate the given spectrahedral shadow, as  the analytical  volume metric for  spectrahedral shadows  remains an open problem  and the containment problem for spectrahedral shadows is co-NP-hard \cite{kellner2013}.  However,  polyhedral approximation of spectrahedral shadows is feasible, and  polyhedrons are readily written as spectrahedrons (see Lemma \ref{propos of H-polyhedron}). Thus, it is practical to use polyhedral approximation to reduce the size and eliminate the  lifted dimensions  for   spectrahedral shadows, on condition that the accuracy and time efficiency meet expectations, 

 Consider a full-dimensional bounded spectrahedral shadow $\mathcal{S}_g = \{x \in \mathbb{R}^{n} \! : \exists \, y\in \mathbb{R} ^m, \, \varLambda(x, y) \succeq 0 \} $.  For an arbitrary   direction $a^i \in \mathbb{R}^{n}$, it is observed that  $\mathcal{H}^i= \{x\in \mathbb{R}^{n} \! : (a^i)^T x \leq b^i \} $     is a half-space externally tangent to  $\mathcal{S}_g$ when $b^i = \max_{x \in \mathcal{S}_g}  (a^i)^T x $. Thus, by choosing $s_r$ directions $a^1, a^2, ..., a^{s_r}  \in \mathbb{R}^{n}$, a polyhedron $\mathcal{P}(A^r, b^r)$ externally tangent to  $\mathcal{S}_g$ is computed by solving the SDP problem
   	\begin{equation} \label{Polyhedral Approximation problem}
 	\begin{aligned}
 	(\bar{x}^{1}\!,...,\bar{x}^{s_r}\!, \bar{y}^1\!,...,  &\,\bar{y}^{s_r}) 	 =  \arg \hspace{-1pt} \max_{x^1\!,...,x^{s_r}\!, y^1\!,...,y^{s_r} } \; \, \sum_{i=1}^{s_r}  (a^i)^T x^i \quad \\ 
 		& \quad s.t., \;   \, \varLambda(x^i, y^i) \succeq 0 , \; \forall \, 1 \leq i \leq s_r ,
 	\end{aligned}
 \end{equation}
 where $x^i\in \mathbb{R}^{n}$, $y^i\in \mathbb{R}^{m}$, $(A^r)^T = [a^1, a^2, ..., a^{s_r}] $ and $b^r =  [(a^1)^T \bar{x}^{1}, (a^2)^T \bar{x}^{2}, ..., (a^{s_r})^T \bar{x}^{s_r}]^T$.  By Lemma \ref{propos of H-polyhedron},  $\mathcal{P}(A^r, b^r)$ is further written as a spectrahedron with the size $s_r$. To make the polyhedral approximation for  $\mathcal{S}_g$ in different directions as uniform as possible, we apply the geometric procedure   in  \cite{lovisolo2001} to  generate the  direction vectors $a^1, a^2, ..., a^{s_r}$, which ensures $a^1, a^2, ..., a^{s_r} \in \mathbb{R}^{n}$ with arbitrary $n$ and $s_r$ are  unit  vectors almost uniformly distributed on an $n$-dimensional hyper-sphere.
  
 The above is a brief introduction to polyhedral approximation for $\mathcal{S}_g$. In  Section \ref{section: Acceleration using  Sparsity }, we   discuss how to  improve the time efficiency  in practice. Moreover,  the effectiveness   will be demonstrated    by numerical examples in Section \ref{section: state estimation examples}.

\subsection{Acceleration using  Sparsity} \label{section: Acceleration using  Sparsity }
When applied to SE, RA, and FD,  spectrahedral shadows with large size and lifted dimensions typically result from the propagation of some basic spectrahedral shadows through  set operations  in Section \ref{section: set operations}, which endows these sets with sparse structures. Specifically, such spectrahedral shadows  $\mathcal{S}_g = \{x \in \mathbb{R}^{n} \! : \exists \, y\in \mathbb{R} ^m, \, \varLambda(x, y) \succeq 0 \} $ with size $s_g$    can be decomposed into
\begin{equation} \label{sparse form of spectrahedral shadows}
	\begin{aligned} 
	\varLambda(x, y) =  \left[ \begin{matrix}
	 	 \varLambda^{(1)}(x, y)     &	\textbf{0} & \cdots  & \textbf{0} \\
	 	\textbf{0}  &	 \varLambda^{(2)}(x, y)  & \cdots  & \textbf{0} \\
	 	\vdots &	\vdots & \ddots  & \vdots\\
	 	\textbf{0}  &	\textbf{0} & \cdots  & \varLambda^{(l)}(x, y)
	 \end{matrix} \right] 
	\end{aligned}
\end{equation}
where $\varLambda^{(i)}(x, y) =  \varLambda^{(i)} +\sum_{j=1}^n{x_jA_j^{(i)}}+\sum_{j=1}^m{y_jB_j^{(i)}}$,  $ \varLambda^{(i)}, A_j^{(i)},  B_j^{(i)} \in \mathbb{S}^{s_g^{(i)}}$, and the inequality $s_g^2 \gg n+m $ typically holds. 

In practice,  the  efficiency of implementing the set validations in Section \ref{section: Set Validations} and the complexity reduction strategies in Section \ref{section: Complexity Reduction Strategy}   is dominated by solving certain SDP problems, e.g.,  \eqref{emptyness check problem}, \eqref{boundary point sdp} and \eqref{Polyhedral Approximation problem}. In the following parts, we will take  \eqref{Polyhedral Approximation problem} as an example to show that through utilizing the structure of $\mathcal{S}_g$,  these SDP problems can be solved  far more efficiently.

When directly solving  \eqref{Polyhedral Approximation problem},  the state-of-the-art interior point method for SDP \cite{jiang2020} and  SDP solvers (e.g., Mosek \cite{mosek2024})  treat the constraints of  \eqref{Polyhedral Approximation problem}  as
\begin{equation} \label{constraints form}
	\begin{aligned} 
X^{(i)} =	\varLambda(x^i, y^i) ,\,  X^{(i)} \succeq 0 , \; \forall \, 1 \leq i \leq s_r
	\end{aligned}
\end{equation}
by introducing  SDP variables  $X^{(1)}, X^{(2)},..,X^{(s_r)} \in \mathbb{S}^{s_g}$. This means that such a problem has $\frac{s_rs_g(s_g+1)}{2}$ equality constraints,  $s_r(m+n)$ scalarized variables, and $ \frac{s_rs_g(s_g+1)}{2}$  scalar  SDP variables. On the other hand, by using the decomposition \eqref{sparse form of spectrahedral shadows}, the constraints in \eqref{Polyhedral Approximation problem} can be written as 
\begin{equation} \label{constraints of sparse form}
	\begin{aligned} 
	\varLambda^{(i)}(x^j, y^j) \succeq 0 , \; \forall \, 1 \leq i \leq l ,\, 1 \leq j \leq s_r.
	\end{aligned}
\end{equation}
With   \eqref{constraints of sparse form},  we further consider  the dual problem of \eqref{Polyhedral Approximation problem}, i.e.,
   	\begin{equation} \label{dual of Polyhedral Approximation problem}
	\begin{aligned}
	&  \quad	(\bar{Z}^{(1,1)}\!,..., \,\bar{Z}^{(s_r,l)}) =  \arg \min_{Z^{(i,j)}} \; \, \sum_{i=1}^{s_r}  \sum_{j=1}^{l}  \mathrm{tr}(\varLambda^{(j)}Z^{(i,j)}) \quad \\ 
	& \quad s.t., \;    Z^{(i,j)} \succeq 0 , \; \forall \, 1 \leq i \leq s_r , \,  1 \leq j \leq l, \\
		& \qquad  \quad a^i_k +  \sum_{j=1}^{l}  \mathrm{tr}(A_k^{(j)}Z^{(i,j)}) = 0 , \;  \forall \, 1 \leq i \leq s_r , \,  1 \leq k \leq n, \\
		& \qquad  \quad   \sum_{j=1}^{l}  \mathrm{tr}(B_t^{(j)}Z^{(i,j)}) = 0 , \;  \forall \, 1 \leq i \leq s_r , \,  1 \leq t \leq m,
	\end{aligned}
\end{equation}
where the multiplier  $Z^{(i,j)} \in \mathbb{S}^{s_g^{(j)}}$. Note that since  $\mathcal{S}_g$ is full-dimensional, there exists a  congruence transformation such that the inequalities \eqref{constraints of sparse form} are strictly feasible \cite[Corollary 5]{ramana1995}. By Slater condition, the strong duality holds for \eqref{Polyhedral Approximation problem} and \eqref{dual of Polyhedral Approximation problem}, and hence $(a^i)^T \bar{x}^{i} =  \sum_{j=1}^{l}  \mathrm{tr}(\varLambda^{(j)}\bar{Z}^{(i,j)})$. 

The problem \eqref{dual of Polyhedral Approximation problem} has $s_r(m+n)$ equality constraints  and $ \sum_{i=1}^{l}\frac{s_r s_g^{(i)}(s_g^{(i)}+1)}{2}$  scalarized  SDP variables. Due to  $s_g^2 \gg n+m $ and $s_g = \sum_{i=1}^{l} s_g^{(i)}$,  both the equality constraints and  scalarized  SDP variables have been significantly  reduced compared with the original problem \eqref{Polyhedral Approximation problem}. We give the following example to  illustrate the  improvement of efficiency. 

 \begin{Exam} \label{example of polyhedron approximation}
	Consider the  following dynamics commonly seen in SE, RA and FD
   \begin{equation} \label{Set-based dynamics}
		\begin{aligned}
		\mathcal{X}_{k+1}  = \mathcal{A}_k\circ \mathcal{X}_{k} \oplus \mathcal{B}_k \circ \mathcal{U}_{k},
		\end{aligned}
	\end{equation}
	where $\mathcal{A}_k,\mathcal{B}_k\in \mathbb{R}^{n \times n} $ and $\mathcal{X}_{k},\mathcal{U}_{k} \subset \mathbb{R}^{n} $. Assume that $X_0$ is an  H-polyhedron with $n_h$ hyperplanes  and each  $\mathcal{U}_{k}$ is an ellipsoid. By representing $\mathcal{X}_0$ and $\mathcal{U}_{k}$ as a spectrahedral shadow, $\mathcal{X}_{k} $  will be a spectrahedral shadow with large size and lifted dimensions if $k$ is large enough. For instance, if $n$ = 3, $n_h = 5$,  the size and lifted dimension of $\mathcal{X}_{30}$ are $s_g= 126$ and $m_g = 90$. If we compute a size-reduced spectrahedron $\hat{\mathcal{X}}_{30}$  with the size $s_r = 20$ using the  polyhedral approximation  in Section \ref{section: Polyhedral Approximation}, the  problem \eqref{dual of Polyhedral Approximation problem} will have $n_{ec} = 157500$ equality constraints,  $n_{sv} = 1860$ scalar variables and $n_{sdp} = 157500$ scalarized  SDP variables. As a comparison,  $\mathcal{X}_{30}$ can be decomposed into the form of \eqref{sparse form of spectrahedral shadows} with $l=36$, and  the  problem \eqref{Polyhedral Approximation problem} will only have $n_{ec} =1860$ equality constraints and $n_{sdp} = 6100$ scalarized  SDP variables.  Moreover, Table \ref{Average time consumpition} shows the  scale and solving time  for the  problems \eqref{Polyhedral Approximation problem} and \eqref{dual of Polyhedral Approximation problem} with different $n+m_g$, $s_g$ and $s_r$, where  solving time $t_{a}$ is averaged over 100 runs\footnote{All numerical examples in this paper are carried out in MATLAB 2024b on an AMD R9 7940HX processor with 2$\times$16 GB memory, and  the formulated optimization problems are solved by Mosek 10.2.8 \cite{mosek2024}.}. From Table \ref{Average time consumpition}, it is observed that the polyhedral approximation  in Section \ref{section: Polyhedral Approximation} is remarkably accelerated by   utilizing the sparsity of $\mathcal{X}_k$. With fixed $s_r$,  the time consumption to solve the problem \eqref{dual of Polyhedral Approximation problem} increases linearly in terms of $n+m_g$ and $s_g$. Moreover, the problem \eqref{Polyhedral Approximation problem} becomes  unsolvable when $n+m_g = 150$ and $s_g = 210$ due to out of memory, but the problem \eqref{dual of Polyhedral Approximation problem} still can be efficiently solved.
\end{Exam}

  \begin{table}[!htbp]\footnotesize 
	\centering
	\renewcommand{\arraystretch}{1.08}
	\setlength{\tabcolsep}{4.7pt} 
	\caption{ Problem scale  and  average solving time.}
	\label{Average time consumpition}
	\begin{tabular}{lllllllll}
		\toprule
		& &  & \multicolumn{3}{c}{Problem \eqref{Polyhedral Approximation problem}}  & \multicolumn{3}{c}{Problem \eqref{dual of Polyhedral Approximation problem}}   \\  \cmidrule(lrr){4-6} \cmidrule(lr){7-9} 
		$n+m_g {}^{1}$& $s_g$ & $s_r$ & \, $n_{ec}$ & \,$n_{sdp}$ & \,$t_{a}$\,[s] & \, $n_{ec}$ & \,$n_{sdp}$ & \,$t_{a}$\,[s] \\ \midrule
		\multirow{2}{*}{\;50} & \multirow{2}{*}{70} &  10 \!\!\!\!  & 24850 & 24850\!\!\!\! & 5.08 & 500\hspace{-5pt} & 1600 & 0.05 \\
		&  & 15   & 37275 & 37275 & 7.44 & 750 &  2400 & 0.07\\ \cmidrule(lr){1-9} 
			\multirow{2}{*}{\;100}& \multirow{2}{*}{140} & 10   & 98700 & 98700 & 117 & 1000 & 3200 & 0.08 \\
		&  & 15 & 148050   & 148050 & 165 &1500 & 4800 &0.11 \\ \cmidrule(lr){1-9} 
			\multirow{2}{*}{\;150}& \multirow{2}{*}{210} & 10 & 221550 &221550 &  ----- &1500 &4800 &0.12 \\
		&  & 20 & 443100 & 443100 & ----- &3000 &9600 &0.23\\
		\bottomrule
	\end{tabular}
	\begin{minipage}{\linewidth}
		 \raggedright
		\vspace{2pt}${ }^1$\,Note that the complexity of the problems \eqref{Polyhedral Approximation problem} and \eqref{dual of Polyhedral Approximation problem} is  proportional to $n+m_g$, instead of the single $n$ or $m_g$.
	\end{minipage}
	
\end{table}

\section{State Estimation with Mixed Polytopic and Quadratic Uncertainties }   \label{section: state estimation examples}

This section considers  set-membership estimation \cite{alamo2005} to demonstrate   set operations  of spectrahedral shadows   and their space overhead  and efficiency advantages  for high-dimensional systems.  Specifically, consider the linear system 
   \begin{equation} \label{system model}
	\begin{aligned}
		x_{k+1}  = \mathcal{A}  x_{k} + \mathcal{B} u_k + \mathcal{L}\omega_k, \quad		y_{k}  = \mathcal{C} x_{k}  +\mathcal{F} v_k 
	\end{aligned}
\end{equation}
with state $x_k \in \mathbb{R}^{n_x} $, input $u_k \in \mathbb{R}^{n_u}$, output $y_k\in \mathbb{R}^{n_y}$, disturbance $\omega_k\in \mathbb{R}^{n_{\omega}}$ and measurement noise $v_k\in \mathbb{R}^{n_v}$. This example considers the case that $x_0$ and   $\omega_k$ are bounded by component, and the energy of   $v_k$ is  limited. Specifically, $x_0 \in \mathcal{X}_0$, $\omega_k \in \mathcal{W}$ and  $v_k \in \mathcal{V}$, where $\mathcal{V}$ is an ellipsoid, and both $\mathcal{X}_0$ and $\mathcal{W}$ are hypercubes. 

Set-membership estimation is a classical set-based state estimator  proven to be  optimal  for linear systems \cite{cong2021}, which can produce the tightest state estimation set $X_k$ containing $x_k$ if all computation is  exact. For   set-membership estimation,  $\mathcal{X}_k$ is obtained from   $\mathcal{X}_k = \bar{\mathcal{X}}_{k} \cap \tilde{\mathcal{X}}_{k}$, where 
   \begin{subequations} \label{set-membership estimation}
	\begin{align}
		 \bar{\mathcal{X}}_{k+1} & = (\mathcal{A} \circ \mathcal{X}_{k} + \mathcal{B} u_k) \oplus  \mathcal{L} \circ W, \;\;\\	\tilde{\mathcal{X}}_{k} &=	(y_{k} + (-\mathcal{F} \circ \mathcal{V}) ) \circ C \label{consistent state set}
	\end{align}
\end{subequations}
 According to \eqref{consistent state set}, the consistent state set $\tilde{\mathcal{X}}_{k} $ will be an unbounded set if the number of system sensors is fewer than system states (i.e., $ \mathrm{rank}(\mathcal{C})< n_x $).  By representing $X_0$, $\mathcal{W}$ and $\mathcal{V}$ as spectrahedral shadows, we can compute the exact $X_k$ using the set operations in Section \ref{section: set operations}.

 For comparison, we   also compute $\mathcal{X}_k$  using ellipsotopes  and CCGs   (the representation and implementation of ellipsotopes and CCGs are consistent for this task) via the open source code  \cite{kousik2023} and the state-of-the-art complexity reduction strategies of CCGs \cite[Algorithms 1 and 2]{rego2025}. Note that we do not evaluate  set representations like zonotopes and ellipsoids, since the uncertainties  considered in this example are mixed polytopic and quadratic, and hence these set representations require outer approximation to represent $\mathcal{W}$ or $\mathcal{V}$, resulting in extra loss of accuracy. For this case, the example in \cite[Section \uppercase\expandafter{\romannumeral5}-C]{kousik2023} has shown that  ellipsotopes and CCGs maintain comparable speed but provide a tighter result.
 
   \begin{figure}[!t]
 	\centering
 	\subfigure{
 		\hspace{0pt} \includegraphics[scale=0.262]{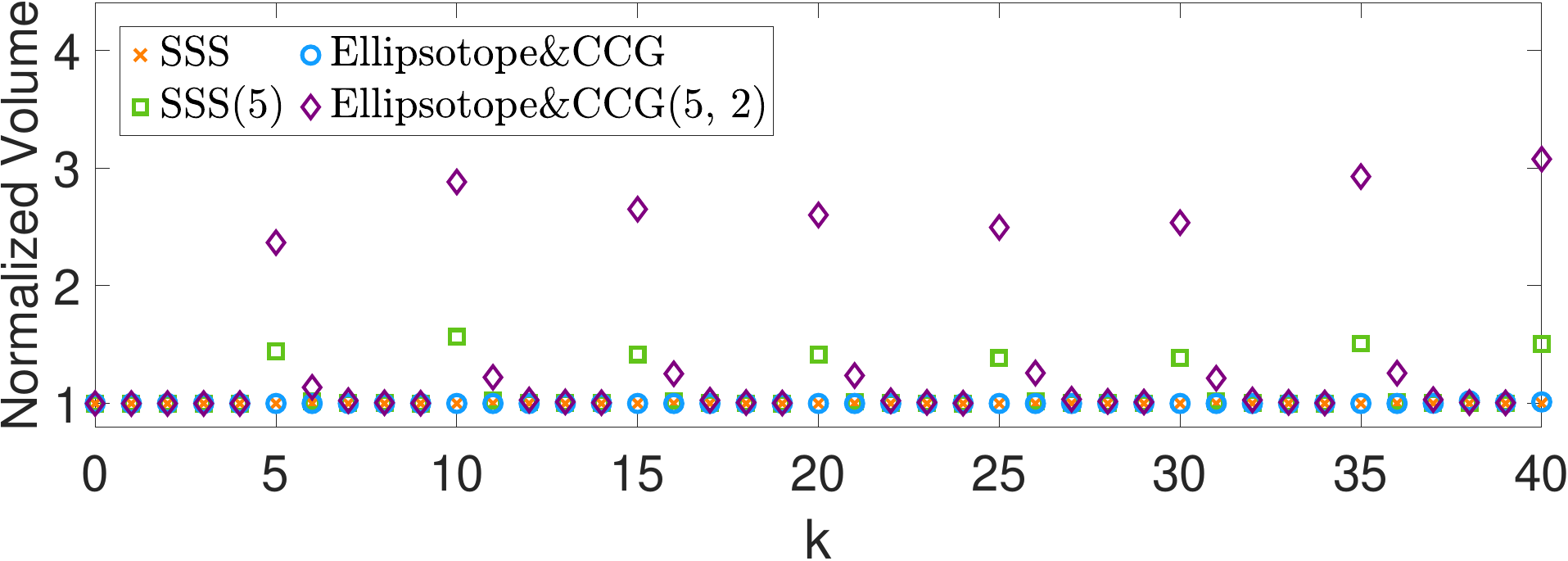}
 	}
 	\subfigure{
 		\hspace{-4pt} \includegraphics[scale=0.262]{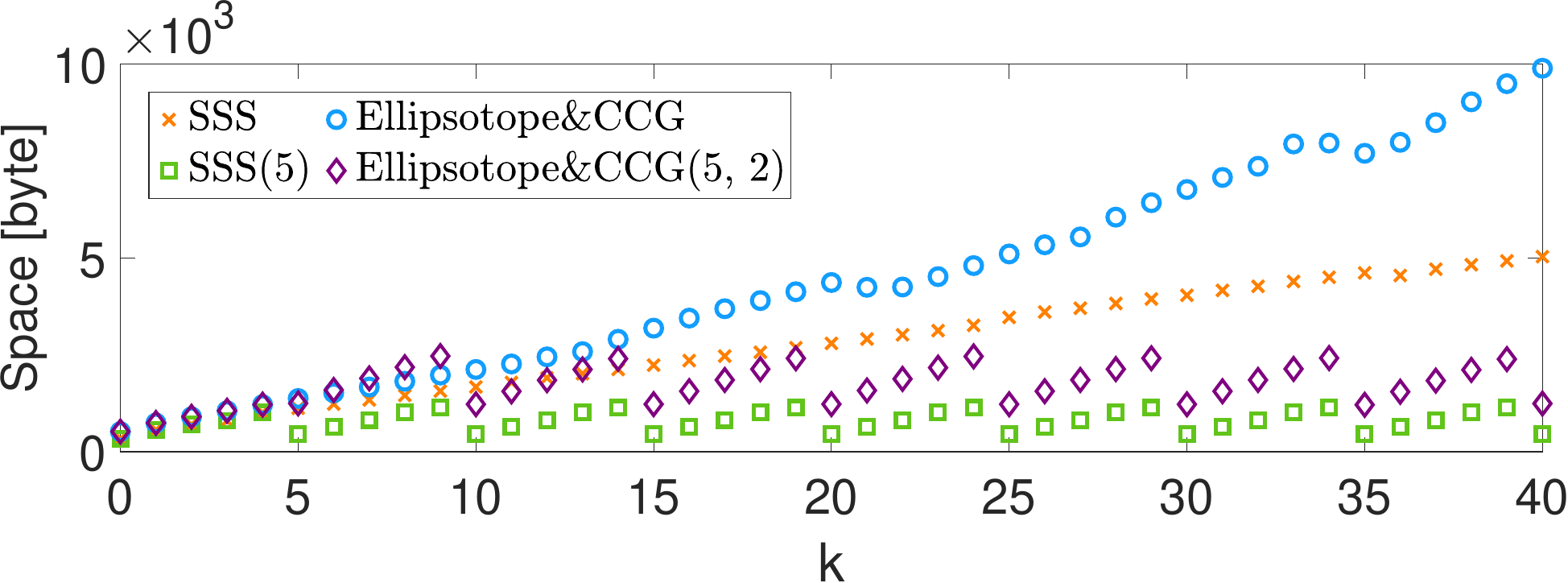}
 	}
 	\vspace{-4pt}
 	\caption{The average volume and space overhead of $X_k$ using $\textit{SSS}$, $\textit{SSS(5)}$, $\textit{Ellipsotope\&CCG}$ and $\textit{Ellipsotope\&CCG(5,2)}$ on random  2-dimensional systems.} \label{fig8}
 \end{figure}
 
 We first evaluate  $\mathcal{X}_k$ for $0\leq k \leq 40$  on 100 random linear systems \eqref{system model} with the dimension $d = n_x =n_u = n_y = n_{\omega} = n_v =2$. As shown in  Fig.\ref{fig8}, $\textit{SSS}$ denotes the results using  spectrahedral shadows  without reducing complexity,   $\textit{Ellipsotope\&CCG}$ denotes  the results using  ellipsotopes (or equivalently, CCGs) without reducing complexity, $\textit{SSS}(s_r/n_x)$ denotes the results using spectrahedral shadows and applying the complexity reduction strategy  in Section \ref{section: Polyhedral Approximation} per 5 steps that approximates $\mathcal{X}_k$ as an $n_x$-dimensional spectrahedron with size $s_r$,  and $\textit{Ellipsotope\&CCG}(n_g/n_x,\, n_c/n_x)$ denotes the results using ellipsotopes and applying the complexity reduction strategy in \cite[Algorithms 1 and 2]{rego2025}  per 5 steps that approximates $\mathcal{X}_k$ as an $n_x$-dimensional ellipsotope with  $n_g$ generators and $n_c$ constraints. The parameter $d_i$ in  \cite[Algorithm 1]{rego2025} is set as $ \frac{1}{n_p}$  according to the recommendation in \cite{rego2025}. The methods $\textit{SSS}$ and $\textit{SSS}(s_r/n_x)$ use sparse forms (see Remark \ref{sparse form remark}). Thus, for a fair comparison, all parametric matrices of ellipsotopes and CCGs that possibly have a large number of zero elements  (e.g., constraints and generator matrices)  are compressed into sparse matrices  when calculating the space overhead. The volume  results  are normalized based on $\textit{SSS}$,  as  $X_k$  calculated by $\textit{SSS}$ is the \textit{exact} state estimation set.

 Apart from  results in Fig.\ref{fig8}, the average time consumption in each step is  0.37\,ms ($\textit{SSS}$), 0.31\,ms ($\textit{Ellipsotope\&CCG}$), 4.61\,ms ($\textit{SSS}(5)$) and 2.17\,ms ($\textit{Ellipsotope\&CCG}(5,2)$). From these results, it can be seen that the space overhead of set representations will increase constantly if no complexity reduction strategy is implemented, but reducing the complexity of  set representations  results in a loss of accuracy and extra time overhead. Compared with ellipsotopes and CCGs, spectrahedral shadows have smaller space overhead, especially for high-order sets. The time consumption of  $\textit{SSS}$ and $\textit{Ellipsotope\&CCG}$ is   comparable,  but the time consumption of $\textit{SSS}(5)$  is about 2.1 times  that of $\textit{Ellipsotope\&CCG}(5,2)$ for 2-dimensional cases, and the reason will be discussed later. In Fig. \ref{fig8}, the volume results of $\textit{SSS}(5)$  illustrate the effectiveness of the complexity reduction strategy  in Section \ref{section: Polyhedral Approximation}.
 
  \begin{figure}[!t]
 	\centering
 	\subfigure{
 		\hspace{-6pt} \includegraphics[scale=0.27]{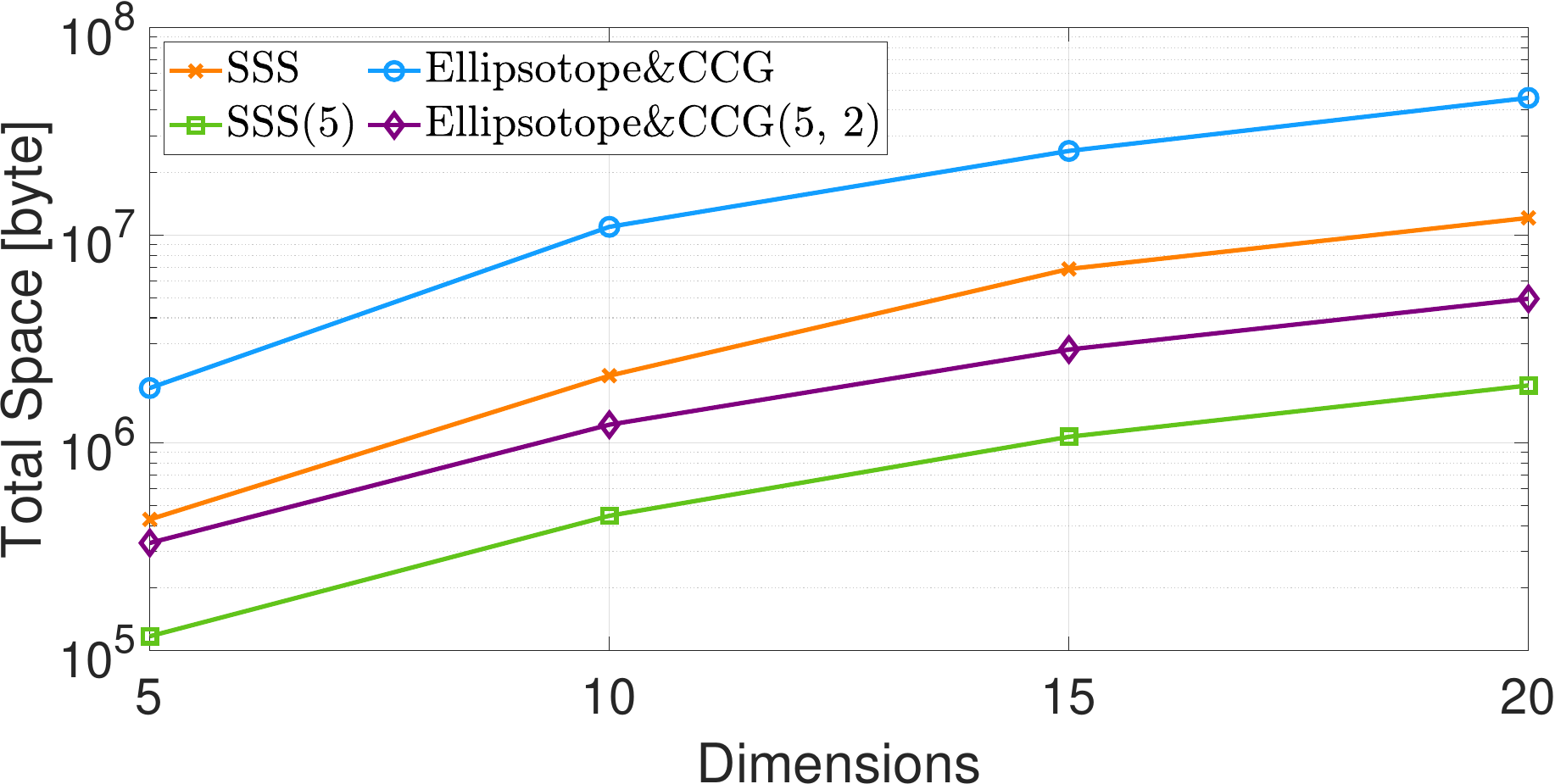}
 	}

 	\subfigure{
 		\hspace{-4pt} \includegraphics[scale=0.262]{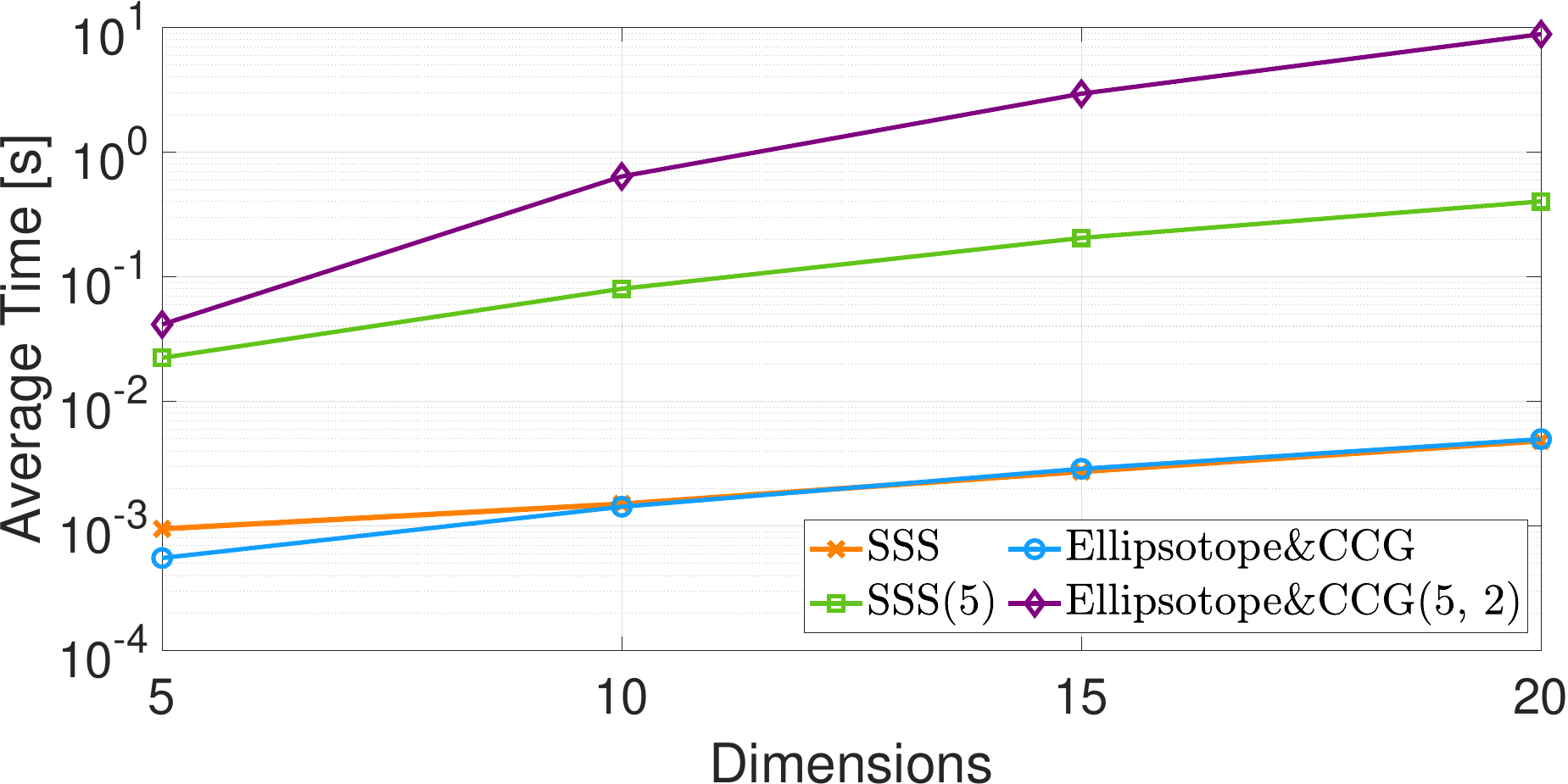}
 	}
 	\vspace{-4pt}
 	\caption{The total space overhead of $X_k$ from $k=0$ to $k=40$  and the average time consumption in each step  on high dimensional systems.} \label{fig9}
 \end{figure}

  Furthermore, we evaluate  $\mathcal{X}_k$ for $0\leq k \leq 40$  with the dimension $d = n_x =n_u = n_y = n_{\omega} = n_v = \{5, \, 10, \, 15, \, 20 \}$. The results are averaged  over 100 random systems \eqref{system model} and shown in Fig. \ref{fig9}. Consistent with research like \cite{kochdumper2020}, the volume results    were not considered  for this high dimensional example, since we aim at illustrating the space overhead and efficiency advantages  of  spectrahedral shadows,  and calculating the exact volume  of high-dimensional spectrahedral shadows is impractical due to its NP-hardness$^{2}$. 
  
  From Fig. \ref{fig9}, it is  observed that the space overhead of spectrahedral shadows is remarkably lower than that of ellipsotopes and CCGs, regardless of whether the complexity reduction strategy is adopted. Although the computational complexity of operations like linear mapping and Minkowski sum of  spectrahedral shadows is higher than that of ellipsotopes and CCGs (see Section \ref{section: set operations} and \cite[Section \uppercase\expandafter{\romannumeral4}-B]{kousik2023}), the time consumption of  $\textit{SSS}$ and $\textit{Ellipsotope\&CCG}$ is still comparable on high dimensional cases. This is because the practical time efficiency is determined by both the computational complexity and the input scale, and the  time consumption caused by the high computational complexity of  spectrahedral shadows is offset by the time savings brought by their low space overhead.  Different from the 2-dimensional case, the time consumption of  $\textit{SSS}(5)$   is lower than that of  $\textit{Ellipsotope\&CCG}(5,2)$ for high dimensional cases, and the difference becomes more noticeable with the increase in dimensions. This is because the time consumption of $\textit{Ellipsotope\&CCG}(5,2)$  mainly results from the constraint reduction strategy   \cite[Algorithm 2]{rego2025} that has $\mathcal{O}(n_{c}^r n_c^2n_g^2)$ complexity, where $n_{c}^r$, $n_c$ and $n_g$ are the number of  constraints  to be reduced, the number of constraints, and the number of generators, respectively. However, with the sparse acceleration strategy in Section \ref{section: Acceleration using  Sparsity }, the time consumption of the polyhedral approximation in Section \ref{section: Polyhedral Approximation}  increases linearly in terms of the dimension, the lifted dimension and the size of spectrahedral shadows,  just as demonstrated in Example \ref{example of polyhedron approximation}. In summary, the above results  demonstrate the space overhead  and efficiency advantages of spectrahedral shadows for modeling uncertainties with complex form (e.g., mixed polytopic and ellipsoidal), especially for high dimensional systems.
 
 \section{Reachable set of Minkowski-Firey $\textit{L}_\textit{p}$ sums of ellipsoids}
 This section uses  the example of reachable analysis taken from \cite{halder2020}  to demonstrate the Minkowski-Firey $\textit{L}_\textit{p}$ sum of spectrahedral shadows. Specifically, consider the linear system with the dynamics $	x_{k+1}  = \mathcal{A}  x_{k} + \mathcal{B} u_k$, where the initial state $x_0 \in \mathcal{X}_0$ and input $u_k \in \mathcal{U}_k$.    We want to compute a reachable set $X_k$ such that $x_k \in X_k$ holds for arbitrary $x_0$ and $u_k$.   To make the description of uncertainty more flexible, the sets $\mathcal{X}_0$ and  $ \mathcal{U}_k$  in \cite{halder2020} are modeled as  the Minkowski-Firey $\textit{L}_\textit{p}$ sums of ellipsoids, i.e., 
  \begin{subequations} 
  	\begin{align}
  		\mathcal{X}_{0} & = \mathcal{E}(\bar{x}_0,\, Q_{1}^0) +_{p_1} .... +_{p_1} \mathcal{E}(\bar{x}_0,\, Q_{m_x}^0)\;\;\\	\mathcal{U}_{k} &=	 \mathcal{E}(\bar{u}_k,\, U_{1}^k) +_{p_2} ... +_{p_2}  \mathcal{E}(\bar{u}_k,\, U_{m_u}^k),
  	\end{align}
  \end{subequations}
  where $p_1, p_2 \geq 1$, $Q_{i}^0 \in \mathbb{S}^{n_x}$ for $ 1 \leq i \leq m_x$, $U_{j}^k \in \mathbb{S}^{n_u}$ for $ 1 \leq j \leq m_u$, and $\bar{x}_0 \in \mathbb{R}^{n_x}$ and $\bar{u}_k \in \mathbb{R}^{n_u}$ are the nominal initial state and control, respectively.
   \begin{figure*}[!t]
  	\centering
  	\subfigure{
  		\hspace{-5pt} \includegraphics[scale=0.5]{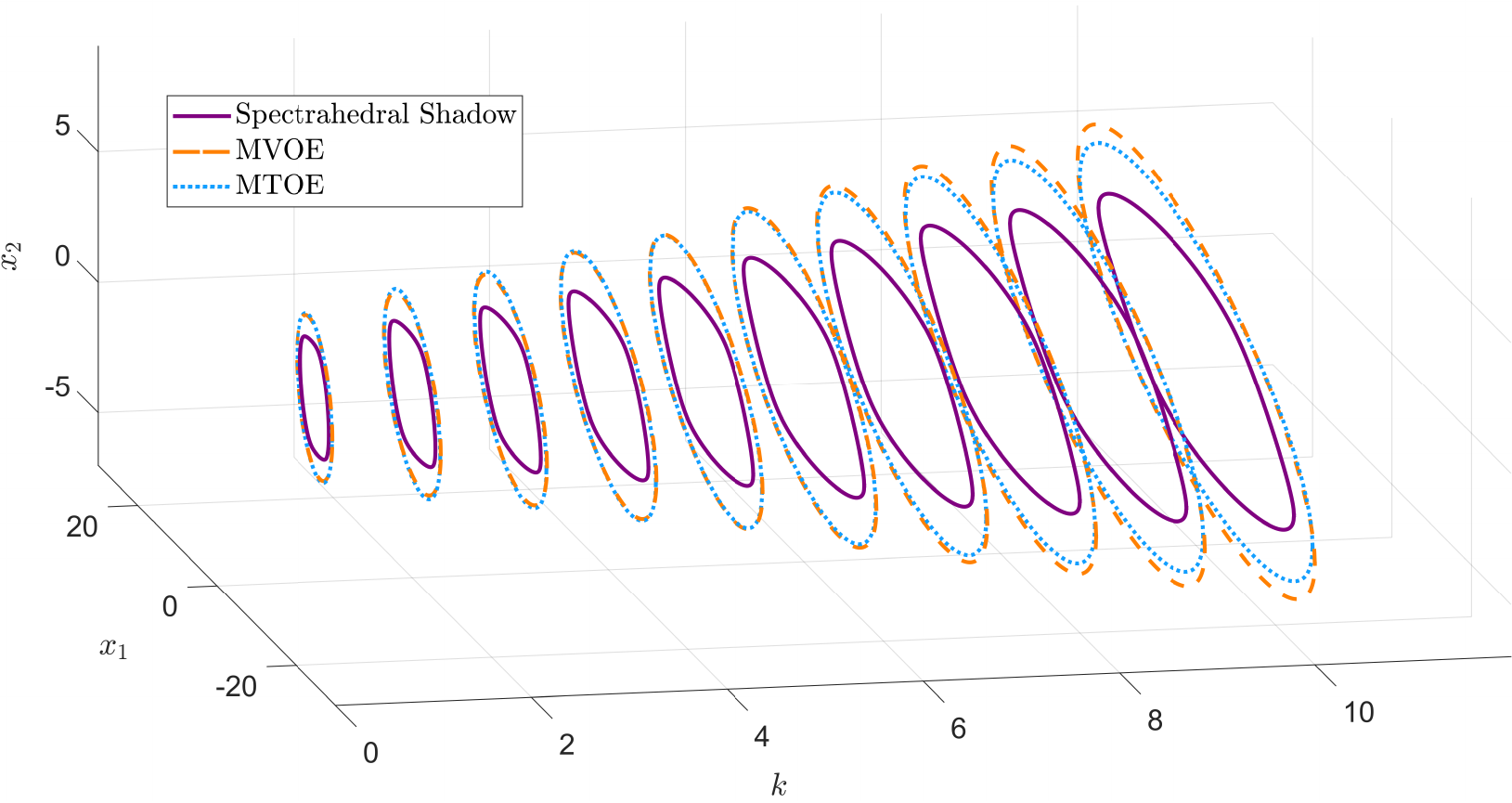}
  	}
  	\vspace{2pt}
  	\caption{The boundaries of reachable sets calculated by spectrahedral shadows, MVOE   and MTOE  for the system given in \cite[Section \uppercase\expandafter{\romannumeral6}-B]{halder2020}.}  	\label{fig10}
  \end{figure*}
  
  \begin{table*}[!htbp]\footnotesize 
  	\centering
  	\renewcommand{\arraystretch}{1.08}
  	\setlength{\tabcolsep}{5.5pt} 
  	\caption{ The volume of reachable sets  and  the average time consumption in each time instant.}
  	\label{results of p sum}
  	\begin{tabular}{llllllllllllll}
  		\toprule
  		&  &\multicolumn{9}{c}{$\mathrm{Vol}(X_k)$}  &   \\  \cmidrule(lrr){2-11}  
  		Methods & $k=1$ & $k=2$ &  $k=3$ & $k=4$ & $k=5$ & $k=6$ &  $k=7$ & $k=8$ & $k=9$ & $k=10$ & Average time [s]  \\ \midrule
  		Spectrahedral shadow & 24.021 &  41.552   & 56.444 & 78.738 & 96.420 & 133.429 & 165.061 & 192.471 &222.664 & 257.184  & 9.383$\times10^{-4}$ \\
  		\cmidrule(lr){1-12} 
  		MTOE \cite[Algorithm 1]{halder2020} &  38.933 &  70.919   & 95.440 & 135.109 & 164.383 & 228.112 & 279.374 & 320.440 &365.306 &416.895 & 5.612$\times10^{-4}$ \\
  		\cmidrule(lr){1-12} 
  		MVOE \cite[Algorithm 2]{halder2020} & 38.984  & 70.202 & 95.239 & 135.565 & 167.269 &234.810 &291.711 &338.986 &391.206 & 451.066&  6.968$\times10^{-4}$ \\
  		\bottomrule
  	\end{tabular}
  \end{table*}	
  
  In this example, we take $p_1 = 3$ and  $p_2 = 1.5$ to compute the reachable set $X_k$ using  spectrahedral shadows.   For this task,   Algorithms 1 and 2 in \cite{halder2020} are the only two available methods in the literature to the best of our knowledge. In this section, both  methods are used for comparison and marked as MTOE and MVOE, respectively. The  error tolerance of \cite[Algorithm 2]{halder2020} is set to $10^{-5}$.  The method to compute $X_k$,  the initial  conditions (e.g., $m_x$ and $m_u$) and the system parameters are all consistent with the example  in \cite[Section \uppercase\expandafter{\romannumeral6}-B]{halder2020}. The results from $k=1,2,..,10$ are shown in Fig. \ref{fig10} and Table. \ref{results of p sum}. It is observed from Fig. \ref{fig10} and Table. \ref{results of p sum} that spectrahedral shadows maintain comparable computational speed but provide much  tighter reachable sets compared with MTOE  and MVOE. This is because the reachable sets computed by spectrahedral shadows are  \textit{exact}, rather than being the outer-approximated ellipsoids  like previous methods.

\section{Conclusion}
The paper investigates the necessary techniques to apply spectrahedral shadows on tasks such as reachability analysis and set-based state estimation.  Spectrahedral shadows can flexibly model  uncertainties in practical tasks, such as those in mixed polytopic and ellipsoidal form, or even convex polynomial form. Compared with the advanced set representations like ellipsotopes and constrained convex generators, spectrahedral shadows support more set operations without approximation, and have lower space overhead  due to their conciseness and sparse structure. Although the theoretical  complexity of common set operations like linear map and Minkowski sum for spectrahedral shadows is higher, the numerical example demonstrates that the practical time consumption of these operations is comparable with ellipsotopes because of the time savings resulting from the lower space overhead. Future work will focus on the flexible complexity reduction methods for spectrahedral shadows to achieve less conservative results.

\appendices

\section*{Acknowledgment}

The authors would like to thank Prof. Feng Xu for the  useful discussion on the topic of set representation, and  Prof. Graziano Chesi for inspiring the authors to apply LMIs in modeling uncertainties.

\section{The computation of $T^*$} \label{The computation of T}
Consider  a parameterized parallelotope $\mathcal{P}_z = \mathcal{Z}(c,\, T^{-1})$ (i.e., an $n$-dimensional zonotope with $n$ generators)   with $T\succ 0$. Our aim is to find an optimal $(c^*, T^*)$ such that $\mathcal{P}_z^* = \mathcal{Z}(c^*,\, (T^*)^{-1})$ is the minimum volume  parallelotope  containing  $\mathcal{S}_g = \{x \in \mathbb{R}^{n} \! : \varLambda +\sum_{i=1}^n{x_iA_i}\succeq 0 \}$. With such $T^*$, $T^* \circ \mathcal{P}_z^* = \mathcal{Z}(T^*c^*,\, I)$ is an $n$-dimensional hypercube. Due to the affine invariance of  minimum volume enclosing sets (see \cite[Section 8.4.3]{boyd2004}), the hypercube  $T^* \circ \mathcal{P}_z^* $ is  the minimum volume  parallelotope containing  $\mathcal{S}_g^{'}= T^*   \circ  \mathcal{S}_g$. Hence, $\mathcal{S}_g^{'}  $  approximates  an isotropic set.

Then we elaborate on the computation of $T^*$. According to Proposition \ref{propos of zonotope}, $\mathcal{P}_z = \mathcal{Z}(c,\, T^{-1})$ can be rewritten as the spectrahedron  $ \langle \varLambda^z , \, \{ A_i^z \} _{i=1}^{n}  \rangle  $ with 
	\begin{equation} \notag
	\begin{aligned}
		&\varLambda^z =  I_{2n} - \mathrm{diag}(c^{\prime}_1,\,c^{\prime}_2,..., c^{\prime}_n,- c^{\prime}_1, -c^{\prime}_2,..., -c^{\prime}_n ),\\
		&A_i^z = \mathrm{diag}(T_{[1, \, i]},\, T_{[2, \, i]} ,..., T_{[n, \, i]},\,  -T_{[1, \, i]},\, -T_{[2, \, i]} ,...,\\
		& \qquad \quad \; \, -T_{[n, \, i]} ),\, \forall \, 1\leq i \leq n,
	\end{aligned}
\end{equation}
where $c^{\prime} = T c$. It is observed that $\mathcal{P}_z$ is also a polytope. According to \cite[Theorem 4.8]{kellner2013}, \cite[Lemma 4.5]{kellner2013} and \cite[Proposition 4.2]{helton2013},   $\mathcal{S}_g \subset \mathcal{P}_z$  iff there exist $O^1, O^2,..., O^{2n} \in  \mathbb{S}^{s_g}$ such that $O^1, O^2,..., O^{2n} \succeq 0$  and
\begin{equation} \label{contain cons}
	\begin{aligned}
		&T_{[i,j]} = \sum_{k=1}^{s_g}\sum_{l=1}^{s_g}A_{j,[k,l]} \, O^i_{[k,l]}, \;  1+c^{\prime}_i =  \sum_{k=1}^{s_g}\sum_{l=1}^{s_g}\varLambda_{[k,l]} \, O^i_{[k,l]}, \\
	-	& T_{[i,j]} = \sum_{k=1}^{s_g}\sum_{l=1}^{s_g}A_{j,[k,l]} \, O^{i+n}_{[k,l]}, \; 1-c^{\prime}_i =  \sum_{k=1}^{s_g}\sum_{l=1}^{s_g}\varLambda_{[k,l]} \, O^{i+n}_{[k,l]}, \\
	 & 	\qquad \qquad\qquad \qquad \forall\, i =1, .., n, \; j = 1, .., n  \\
	 \noalign{\vskip -10pt}
	\end{aligned}
\end{equation}
where $A_{j,[k,l]}$ denotes the element in $k$-th row  and $l$-th column of   $A_j$. 
Since the  volume of  $\mathcal{P}_z$ is  $ \mathrm{vol}(\mathcal{P}_z) = 2^n \mathrm{det}(T^{-1})$ \cite{alamo2005}, $T^*$ can be obtained by solving the  SDP problem
  	\begin{subequations} \label{Mininum volume parallelotope problem}
	\begin{align}
		&  \min_{T, \, c^{\prime}, \,O^1,...,O^{2n} } \; \, \log \det T^{-1}  \quad \\ 
		& \quad s.t., \;   \, O^1 \succeq 0, ....,  O^{2n} \succeq 0, \; \eqref{contain cons} .
	\end{align}
\end{subequations}

\bibliography{cited}
\bibliographystyle{IEEEtran}

\end{document}